\documentclass[lettersize,journal]{IEEEtran}
\usepackage{amsmath,amsfonts}
\usepackage[linesnumbered,ruled,vlined]{algorithm2e}
\usepackage{array}
\usepackage{subfigure}
\usepackage{textcomp}
\usepackage{stfloats}
\usepackage{url}
\usepackage{verbatim}
\usepackage{graphicx}
\usepackage{cite}
\usepackage{braket}
\usepackage{orcidlink}
\usepackage{diagbox}
\usepackage{makecell}
\usepackage {multirow}
\usepackage{fancyhdr}
\hypersetup{
colorlinks=true,
linkcolor=black,
citecolor=black,
urlcolor = black
}
\begin{document}

\title{Building a Hierarchical Architecture and Communication Model for the Quantum Internet}
\author{Binjie He$^{\orcidlink{0000-0003-2373-0391}}$, ~\IEEEmembership{Student Member,~IEEE,} Dong Zhang$^{\orcidlink{0000-0002-6379-0244}}$, ~\IEEEmembership{Member,~IEEE,} \\ Seng W. Loke$^{\orcidlink{0000-0002-5339-9305}}$, ~\IEEEmembership{Member,~IEEE,} Shengrui Lin$^{\orcidlink{0000-0001-5659-431X}}$,  and Luke Lu
\thanks{Manuscript received May 30, 2023; revised November 28, 2023; accepted December 20, 2023. This work was supported by the National Key Research and Development Program of China under Grant 2023YFB2904000 and Grant 2023YFB2904005. (Corresponding author: Dong Zhang)}
\thanks{Binjie He and Shengrui Lin are with the College of Computer and Data Science, Fuzhou University, China (email: hebinjie33@gmail.com; xmlsr@foxmail.com).}
\thanks {Dong Zhang is with the College of Computer and Data Science, Fuzhou University, China (email: zhangdong@fzu.edu.cn).}
\thanks{Seng W. Loke is with the School of Information Technology, Deakin University, Melbourne, Australia (email: seng.loke@deakin.edu.au).}
\thanks{Luke Lu is with the Cisco Systems, Inc. (email: luklu@cisco.com).}
}

\maketitle
\thispagestyle{fancy}
\fancyhead{}
\lhead{\footnotesize{This paper has been published in IEEE Journal on Selected Areas in Communications, vol. 42, no. 7, pp. 1919-1935, July 2024. DOI: 10.1109/JSAC.2024.3380103}  \\ \footnotesize{©IEEE 2024. Personal use of this material is permitted. Permission from IEEE must be obtained for all other uses, in any current or future media including reprinting/republishing this material for advertising or promotional purposes, creating new collective works, for resale or redistribution to servers or lists, or reuse of any copyrighted component of this work in other works.}}
\cfoot{\quad}
\renewcommand{\headrulewidth}{0pt}
\pagestyle{empty}
\begin{abstract}
The research of architecture has tremendous significance in realizing quantum Internet. Although there is not yet a standard quantum Internet architecture, the distributed architecture is one of the possible solutions, which utilizes quantum repeaters or dedicated entanglement sources in a flat structure for entanglement preparation \& distribution. In this paper, we analyze the distributed architecture in detail and demonstrate that it has three limitations: 1) possible high maintenance overhead, 2) possible low-performance entanglement distribution, and 3) unable to support optimal entanglement routing. We design a hierarchical quantum Internet architecture and a communication model to solve the problems above. We also present a W-state Based Centralized Entanglement Preparation \& Distribution (W-state Based CEPD) scheme and a Centralized Entanglement Routing (CER) algorithm within our hierarchical architecture and perform an experimental comparison with other entanglement preparation \& distribution schemes and entanglement routing algorithms within the distributed architecture. The evaluation results show that the entanglement distribution efficiency of hierarchical architecture is 11.5\% higher than that of distributed architecture on average (minimum 3.3\%, maximum 37.3\%), and the entanglement routing performance of hierarchical architecture is much better than that of a distributed architecture according to the fidelity and throughput.
\end{abstract}


\section{Introduction}
In recent decades, quantum physics has revolutionized computer science. Quantum Internet, one of the most eye-catching technologies, is the foundation of distributed quantum computing\cite{cuomo2020towards}. Quantum Internet utilizes quantum repeaters\cite{muralidharan2016optimal} and quantum operations (e.g., quantum teleportation\cite{bennett1993teleporting, bouwmeester1997experimental}, entanglement swapping\cite{olmschenk2009quantum, pan1998experimental}) to transfer qubits between quantum systems.

At present, the quantum Internet is tentatively envisioned as a heterogeneous network, which may consist of sensor networks, satellite quantum networks, and terrestrial quantum Internet\cite{kozlowski2023rfc}. There is no standard architecture for the terrestrial quantum Internet until now, which is still developing only in the draft stage\cite{kozlowski2023rfc}. However, the distributed architecture\cite{van2021quantum, li2021building} is one of the possible solutions, which utilizes quantum repeaters or dedicated entanglement sources in a flat structure for entanglement preparation \& distribution. Many quantum Internet kinds of research are based on a distributed architecture, such as Distributed Entanglement Preparation \& Distribution (DEPD)\cite{ladd2006hybrid, bogdanovic2017design, zou2003generation, barrett2005efficient, campbell2008measurement, jones2016design, yuan2010entangled, aspelmeyer2003long} and Distributed Entanglement Routing (DER)\cite{shi2020concurrent, pant2019routing, van2013path, gyongyosi2017entanglement, chakraborty2019distributed, caleffi2017optimal}. 

Inspired by the notion of Software Defined Network (SDN)\cite{mckeown2008openflow}, this paper focuses on terrestrial quantum Internet and proposes a hierarchical architecture and a communication model. The hierarchical architecture can support optimal entanglement routing, high-performance entanglement preparation \& distribution, error control, and lower maintenance costs. These advantages will benefit distributed quantum computing\cite{cuomo2020towards}, the most important quantum Internet application. We need to note that the hierarchical architecture is not about replacing the network of quantum repeaters, but it can provide a way to organize the quantum repeaters to serve a large quantum Internet. The contributions of this work include five aspects, as described below.

{\bfseries Analyzing the distributed architecture:}  In Sec. \ref{analysis_of_distributed_arch}, we analyze the distributed architecture qualitatively and quantitatively based on the literature and experiments. The analysis results highlight some shortcomings of distributed architecture, including possible high maintenance overhead, possible low-performance entanglement distribution, and inability to support optimal entanglement routing, due to dependence on infrastructure layer devices (e.g., repeaters) for entanglement preparation \& distribution and lack of a unified control plane.

{\bfseries Building a hierarchical architecture:} In Sec. \ref{desgin_of_arch}, we present a three-layer hierarchical quantum Internet architecture and discuss the scalability and robustness. At the top layer, we deploy the central controller to collect the global network state for supporting optimal entanglement routing. At the middle layer, we design the local domain controller for centralized entanglement preparation \& distribution, reducing maintenance overhead and improving efficiency. At the bottom layer, quantum devices are responsible for executing quantum operations. Our hierarchical architecture is not entirely centralized; a local domain controller is responsible for one domain, and a central controller is responsible for a Quantum Local Area Network (Q-LAN) composed of several domains. We envision that the quantum Internet consists of many Q-LANs and central controllers collaborating through East-West communication. This paper focuses on studying what we call South-North communication, which represents protocols involving entities belonging to different layers of the hierarchical architecture. What we call East-West communication, which represents peer-to-peer protocols, such as cooperation between central controllers, can be the future research direction.

{\bfseries Designing a communication model:} In Sec. \ref{model_and_protocols}, we design a communication model from the physical layer to the transport layer for hierarchical architecture. The communication model mainly considers the process of intra-domain and inter-domain communication and error control. The communication process includes communication requests, entanglement preparation, distribution, swapping control, quantum teleportation control, entanglement routing strategy, and resource reservation. As for error control, qubit decoherence can occur in the communication process due to time\cite{schlosshauer2019quantum}. To control errors caused by quantum operation timeouts, we set entanglement distribution timer $t_d$ and entanglement swapping \& teleportation timer $t_{st}$ for time synchronization of quantum operation. Once the central controller detects a timeout, it stops the communication to prevent the error from further damaging the target qubit, thereby preserving the opportunity to retry the communication.

{\bfseries Proposing the W-state Based Centralized Entanglement Preparation \& Distribution (W-state Based CEPD): } We propose the W-state Based CEPD within the hierarchical architecture (Sec. \ref{physical_layer}) and perform an experimental comparison with other DEPD schemes (Sec. \ref{performance_of_entanglement_pd}). The W-state Based CEPD utilizes the local domain controller in the middle layer instead of the repeaters or dedicated entanglement sources in the bottom layer to perform the entanglement distribution of atom qubits. This scheme decouples entanglement preparation \& distribution from infrastructure layers devices (e.g., repeaters) and improves efficiency.

{\bfseries Proposing the Centralized Entanglement Routing (CER):} In order to perform optimal entanglement routing, we present the CER  (Sec. \ref{network_layer}) and perform an experimental comparison with other DER algorithms (Sec. \ref{performance_of_entanglement_routing}). CER evaluates paths by hops and the network state (environmental interference). The success rate of quantum operations reflects environmental interference (e.g., channel noise, dephasing noise, and depolarizing noise). Thus, we represent the network state by entanglement swapping success rate and entanglement preparation \& distribution success rate.

Sec. \ref{background} introduces the background and preliminaries. Sec. \ref{related_work} presents the related work and Sec. \ref{analysis_of_distributed_arch} analyzes the distributed architecture. We describe the hierarchical quantum Internet architecture in Sec. \ref{desgin_of_arch}. Sec. \ref{model_and_protocols} presents the communication model. Sec. \ref{evaluation} shows the evaluation result. In Sec. \ref{conclusion}, we conclude this work.

\section{Background and Preliminaries}
\label{background}
Quantum communication realizes the transmission and processing of quantum information by quantum channels and techniques. In order to realize quantum communication, we need to build quantum networks or even the quantum Internet. The researchers envision a quantum Internet as a network that supports global quantum communications\cite{wehner2018quantum}. Quantum Internet research is still in its infancy; researchers envision it as a heterogeneous network, which may consist of sensor networks, satellite networks, and terrestrial quantum Internet\cite{kozlowski2023rfc}. Our research focuses on the terrestrial quantum Internet. In this section, we will introduce the preliminaries for the work in our article.

{\bfseries Basic resources in quantum communication:} Qubit is the basic unit of a quantum system. A qubit can represent not only 1 and 0 but also the superposition of 1 and 0\cite{cuomo2020towards}. Quantum entanglement refers to the entanglement of multiple qubits through preparation or interaction so that the quantum state of each qubit cannot be described independently of the other qubits\cite{pan1998experimental}. Entanglements have the feature of nonlocality; that is, once the state of one qubit in the entangled pair is changed, it will immediately affect the state of the other qubits, no matter how far apart they are. Because of the special properties of quantum entanglement, it becomes an indispensable resource for quantum communication.

{\bfseries Basic technical concepts in quantum communication:} Entanglement preparation refers to the preparation of entangled pairs by physical methods, for example, UV pulse passing through the nonlinear crystal, which generates two-photon entangled pairs\cite{pan1998experimental}. Entanglement distribution refers to the entanglement source distributing the prepared entanglement to two or more devices through quantum channels\cite{jones2016design}. Entanglement routing refers to finding a path for quantum communication\cite{chakraborty2019distributed}. Entanglement swapping refers to generating entanglement between two distant target devices through appropriate measurement and classical information assistance\cite{olmschenk2009quantum}. Quantum teleportation utilizes entanglement and quantum operation to teleport a target qubit from one site to another without transmitting the particle directly\cite{bennett1993teleporting}. There are also some technologies not covered in this article but may be integrated into the proposed architecture in the future, such as entanglement purification and quantum error correction.

{\bfseries Basic components in the quantum Internet:} Wehner \emph{et al.}\cite{kozlowski2019towards, wehner2018quantum} proposed the basic components of the quantum Internet. The quantum user is the end node in the network, which is the role that ultimately uses quantum information, such as a quantum computer. The quantum repeater is a relay device in the network that can store qubits in memory and perform entanglement swapping (perhaps with entanglement purification and error correction) to overcome the channel noise for long-distance communication. The classical channel transmits classical information, such as network control information or measurement results. The quantum channel transmits qubits, such as photons. There are also some components not covered in this article but may be integrated into the proposed architecture in the future, such as the quantum switch and quantum router.

{\bfseries Protocol stack in quantum communication:} The protocol stack\cite{illiano2022quantum} is divided into the physical, link, network, transport, and application layers. The physical layer protocol is responsible for entanglement preparation and aims to attempt entanglement generation. The link layer protocol is responsible for entanglement distribution, and aims to generate robust entanglement. The network layer protocol is responsible for entanglement routing and swapping and aims to generate long-distance entanglement. The transport layer protocol is responsible for qubit transmission by quantum teleportation. The application layer can deploy quantum programs for computation.

\section{Related Work}
\label{related_work}
Wehner \emph{et al.} proposed the essential components of the quantum Internet\cite{kozlowski2019towards, wehner2018quantum}, which provides a necessary foundation for our work. Kozlowski \emph{et al.} presented the architectural principles for the quantum Internet\cite{kozlowski2023rfc}, which guides us in designing the control plane of our hierarchical architecture. There has been some research on quantum Internet architecture, as discussed below. The cluster-based quantum Internet\cite{li2021building} is a distributed architecture consisting of Q-LANs (made up of master and slave nodes) and a core network (made up of repeaters). Unlike this paper, the cluster-based architecture is a flat structure, and the research mainly focuses on the discussion of the infrastructure layer, and the discussion of the control layer is not enough. Quantum Recursive Network Architecture (QRNA)\cite{van2021quantum, van2011recursive} is a hierarchical distributed quantum Internet that provides a multi-layer system in the abstract sense; some infrastructural layer devices are abstracted into a single device to form an abstraction level to facilitate communication. Unlike QRNA, the hierarchical architecture in this paper is innovative in that it provides a multi-layer system in the physical sense to achieve an independent and unified control plane, allowing the infrastructure layer devices to focus on performing specific quantum operations. We also need to mention the research on satellite quantum Internet\cite{chiti2021towards, picchi2020towards}, which introduced the SDN controller into the quantum Internet to assist entanglement distribution and routing, but they did not explicitly propose a hierarchical architecture design. Moreover, a survey of the quantum Internet protocol stack is presented in\cite{illiano2022quantum}, providing an important reference for designing our communication model.

Cavity Quantum Electrodynamics Based Entanglement Preparation \& Distribution\cite{ladd2006hybrid, bogdanovic2017design, zou2003generation} requires a strict preparation environment in each quantum repeater to prepare and distribute entangled atoms. Double-photon Based Entanglement Preparation \& Distribution\cite{jones2016design, yuan2010entangled, aspelmeyer2003long} requires each quantum repeater to prepare entangled photons for direct distribution, but the photon qubit lifetime is short. The above two kinds of entanglement preparation and distribution methods reflect the long lifetime of atomic qubits and the easy transmission of photon qubits and provide references for us to design entanglement preparation and distribution schemes in hierarchical architectures. A central quantum network node for entanglement distribution is proposed in\cite{avis2023analysis}, utilizing GHZ-states to perform multipartite entanglement distribution. Although the central node is also used in this paper, we utilize the W-state for atom entanglement distribution, which can potentially provide a longer coherence time\cite{heshami2016quantum}. Similarly, the Qonnector is proposed in\cite{yehia2022quantum}, a central node that aims to perform entanglement distribution for a metropolitan network. However, the premise is that the entanglement preparation capability of the Qonnector should be enough to support the terminals of an entire city. Some studies\cite{diadamo2022packet, yoo2021quantum} proposed combining classical information (packet header and tail) and quantum information (payload) to assist in entanglement distribution and to perform packet switching according to information in the header and tail. However, classical information can bring uniquely noise effects on a quantum payload when they are transmitted through the same channel\cite{diadamo2022packet}. Regarding the entanglement distribution protocol, Dahlberg \emph{et al.} proposed the first link layer protocol\cite{dahlberg2019link}. Barrett and Kok offered the Barrett-Kok (BK) protocol to generate entanglement pairs between two remote nodes\cite{barrett2005efficient}. Campbell and Benjamin proposed the Extreme Photon Loss (EPL) protocol to optimize the BK protocol\cite{campbell2008measurement}. MidpointSource\cite{jones2016design} uses a midpoint entangled photon source to distribute entangled photons between quantum repeaters and utilizes teleportation to transfer the quantum information into atom qubits. The above protocol research provides an essential reference for us to design the link layer protocol and analyze the entanglement distribution of a distributed architecture. However, we need to point out that different from the above research focusing on the protocol itself, this paper aims to provide a full-stack communication model in the form of a hierarchical architecture to serve end-to-end communication.

Entanglement routing is one of the vital research areas of the quantum Internet. SLMP\cite{pant2019routing} requires global entanglement distribution before routing and then chooses the shortest path from the nodes with successful entanglement distribution. The global entanglement distribution of SLMP can avoid the nodes with failed entanglement distribution, but it costs a lot of time and resources. Q-PASS and Q-CAST can adapt to arbitrary topology and allow concurrent quantum communication\cite{shi2020concurrent}. Traditional routing algorithms such as Dijkstra, Greedy, and Gradient Routing are also used to select the shortest path in quantum communication\cite{van2013path, gyongyosi2017entanglement, chakraborty2019distributed}. Caleffi proposed Routing Metric to assist routing\cite{caleffi2017optimal}. The above entanglement routing strategies provide material for us to analyze the entanglement routing of distributed architecture. We also select some representative strategies as the performance comparison of CER proposed in this paper.

\section{Analysis of The Distributed Architecture}
\label{analysis_of_distributed_arch}
\begin{figure}[tp]
\centering
\includegraphics[width=3in]{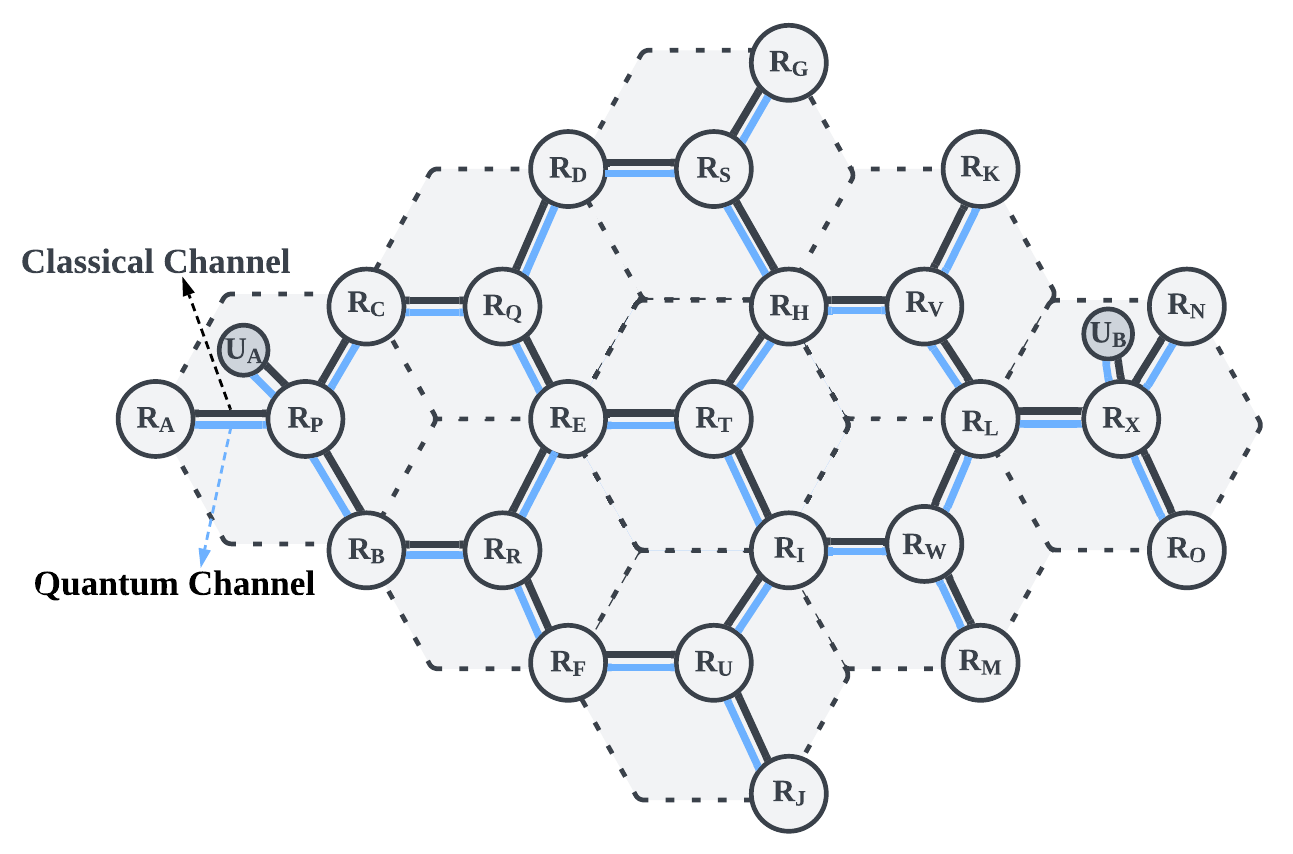}
\caption{An example of a distributed architecture. This distributed architecture example can include users, repeaters, classical and quantum channels. It relies on infrastructure layer devices (can be but not limited to repeaters) to prepare and distribute entanglement pairs for quantum communication.}
\label{distributed_arch}
\end{figure}

The distributed architecture is one possible solution for the terrestrial quantum Internet, possibly part of the heterogeneous quantum Internet in the future. The advantage of a distributed architecture is that it can be modularized to form a network. The distributed architectures can include but are not limited to quantum user, quantum repeater, classical channel, and quantum channel\cite{kozlowski2019towards, wehner2018quantum}. Fig. \ref{distributed_arch} shows an example of a distributed architecture. The core mechanism of the distributed quantum Internet requires infrastructure layer devices (can be but not limited to repeaters) to prepare and distribute entangled pairs to support quantum communication. However, according to our analysis, this mechanism may lead to three problems, as described in the following.

\subsection {Maintenance Cost}
Infrastructure layer devices are responsible for entanglement preparation \& distribution in the distributed architecture. It should be pointed out that in some studies, dedicated entanglement sources are used for entanglement preparation \& distribution. For example, Yehia \emph{et al.} envision a conceptualized Qonnector for entanglement distribution of the entire metropolitan network, but the premise is that the entanglement preparation capability of the Qonnector can support the terminals of an entire city\cite{yehia2022quantum}. In other studies\cite{van2021quantum, van2011recursive}, repeaters are used for entanglement distribution. In either case, the entanglement preparation needs well-designed environments because noise can have a serious impact\cite{nosrati2020robust}.

Presently, the preparation of double-photon entangled pairs is the most used in quantum communication\cite{jones2016design, yuan2010entangled, aspelmeyer2003long}. We can use simple methods to prepare double-photon entangled pairs (e.g., UV pulse passing through the nonlinear crystal\cite{pan1998experimental}). However, the preparation result is random, prompting more resources and operations to filter out the ideal double-photon entangled pairs\cite{PhysRevA.67.030101}. In addition, the short lifetime of photons is not conducive to long-distance quantum communication\cite{takahashi2007high}.

Cavity Quantum Electrodynamics is a popular technique for preparing double-atom entangled pairs to provide qubits with a long lifetime\cite{ladd2006hybrid, bogdanovic2017design, zou2003generation}. However, it usually requires the fabricator to maintain high cavity Q values at liquid-helium temperatures\cite{treussart1998evidence}.

In conclusion, if repeaters are responsible for entanglement preparation, it will be expensive to maintain the vast number of repeaters' preparation environments and resources on the global quantum Internet. A quantitative analysis of maintenance costs is presented in Sec. \ref{scalability_robustness}.

\subsection {Entanglement Distribution}
\begin{figure}[tp]
\centering
\includegraphics[width=3in]{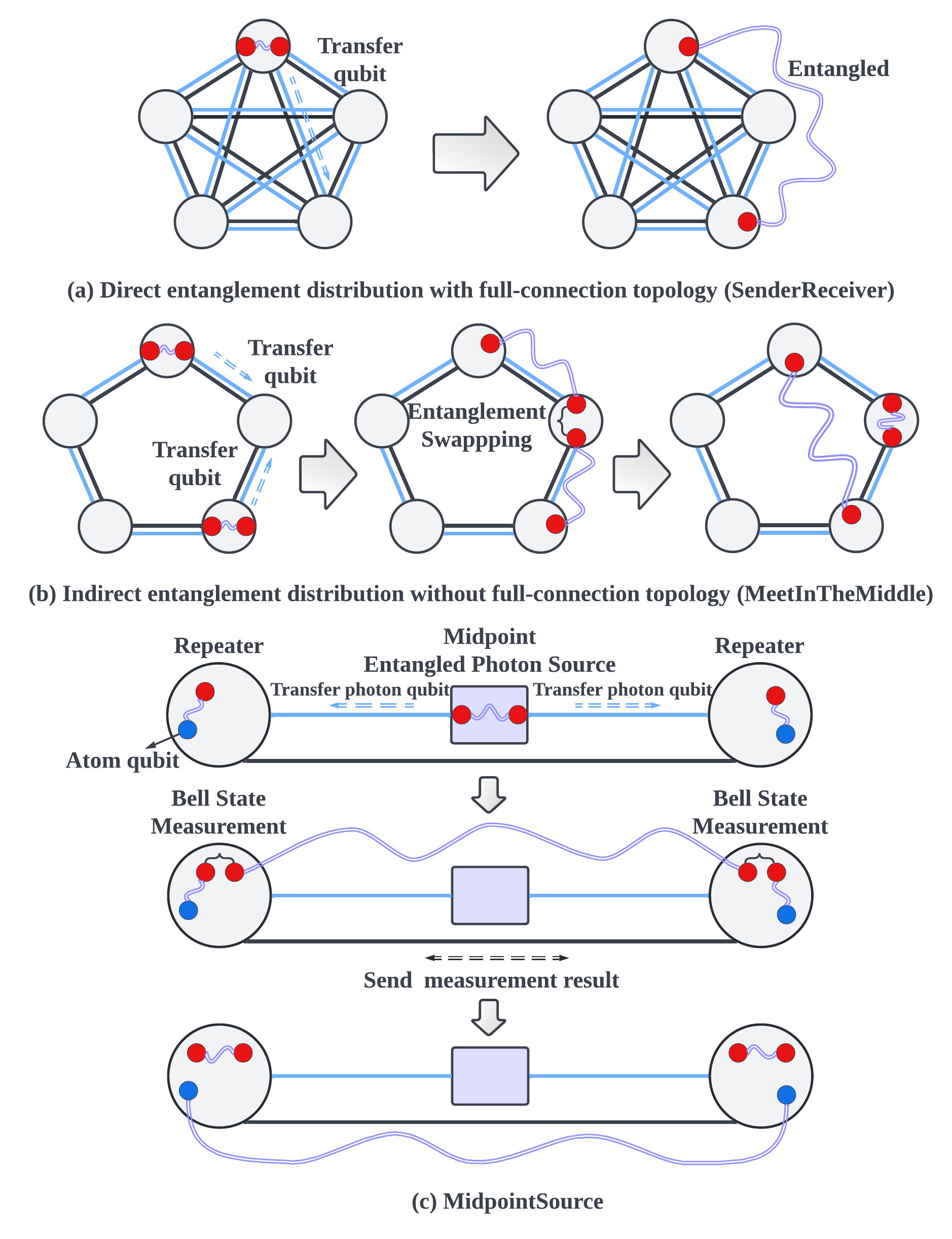}
\caption{Entanglement distribution in distributed architecture. (a) SenderReceiver requires direct entanglement distribution between neighbor repeaters. (b) MeetInTheMiddle requires an intermediate node for performing entanglement swapping to generate entanglement pairs between repeaters. (c) MidpointSource requires a midpoint to distribute atom-atom entanglement between repeaters.}
\label{entanglement_distribution}
\end{figure}

The entanglement distribution in distributed architecture includes three types\cite{jones2016design}:

1) SenderReceiver: It requires one repeater to prepare entanglement for direct entanglement distribution to another repeater, as shown in Fig. \ref{entanglement_distribution}(a). In a network that only uses repeaters for entanglement distribution, a full-connection topology is required if we want to distribute entanglement directly between any two repeaters. However, it is unrealistic to realize the full-connection topology in the area with a large number of devices. We should point out that some studies\cite{diadamo2022packet, yoo2021quantum} suggest that classical information (packet header and tail) and quantum information (payload) can be combined and sent over the same channel. Then, quantum information can be forwarded according to packet header and tail by a Quantum Relay or Quantum Wrapper Switch Router, thereby reducing the number of connections and avoiding the case of full-connection topology. However, classical information can bring uniquely noise effects on a quantum payload when they are transmitted through the same channel\cite{diadamo2022packet}.

2) MeetInTheMiddle: It requires both repeaters to prepare entanglement and transmit the entangled photons to the midpoint for entanglement swapping, as shown in Fig. \ref{entanglement_distribution}(b). However, if two repeaters are close to each other without a quantum channel connection, they still need intermediate nodes for entanglement swapping to complete the distribution. The extra entanglement swapping wastes resources and reduces the efficiency of distribution.

3) MidpointSource: It requires the midpoint to prepare and distribute entangled photons to quantum repeaters. And then, repeaters prepare atom-photon entanglement and perform Bell State Measurement (BSM) to transfer quantum information into atom qubits, as shown in Fig. \ref{entanglement_distribution}(c). MidpointSource is an important step forward from distributed to hierarchical quantum Internet architecture, but it still requires repeaters to prepare atom-photon entanglement, causing maintenance cost problems.

Sec. \ref{performance_of_entanglement_pd} provides the quantitative analysis of entanglement distribution.

\subsection {Entanglement Routing}
Entanglement routing directly determines the success rate of quantum communication. We demonstrate through simulation that for a quantum communication path, the influence of environmental interference on communication quality is greater than that of hops. (see Appendix \ref{environmental_interference_importance}). Therefore, in this study, we assume that environmental interference has a higher priority than hops in the entanglement routing. However, the distributed architecture makes integrating and collecting global environmental interference difficult because it lacks a unified control plane.

At present, the entanglement routing strategy (called DER) used in the distributed architecture has some shortcomings. DER usually selects paths based on hops without considering environmental interference causing non-optimal routing. Non-optimal routing means the qubits transmitted on the chosen path cannot achieve the best fidelity. We can imagine an idealized example that assumes a short path (fewer hops) with very high environmental interference and a long path (more hops) without environmental interference. It is obvious that the fidelity of qubits transmitted through the short path must be lower than that of the long path because of the very high environmental interference. However, DER will still select a short path for communication due to hops, which leads to non-optimal routing. We choose three representative DER to illustrate our claim. Greedy\cite{chakraborty2019distributed} selects the shortest path by the greedy algorithm without considering other factors, so it usually meets the path failure caused by entanglement distribution or swapping failure. SLMP\cite{pant2019routing} performs global entanglement distribution before routing to avoid the distribution failure node, but our experiment points out that global entanglement distribution causes massive resource consumption and time cost (see Table \ref {routing_cost}). Q-Cast\cite{shi2020concurrent} can consider the quantum memory resource to perform entanglement routing, but it still ignores the environmental interference. Sec. \ref{performance_of_entanglement_routing} provides the quantitative analysis of entanglement routing.

\section{Design of Hierarchical Quantum Internet Architecture}
\label{desgin_of_arch}
Based on the analysis in Sec. \ref{analysis_of_distributed_arch}, we conclude that the problems of maintenance and entanglement distribution are caused by entanglement preparation \& distribution relying on infrastructure layer devices (especially repeaters), and the problem of entanglement routing is caused by the distributed architecture lacking a unified control plane. Thus, we propose a three-layer hierarchical quantum Internet architecture to decouple entanglement preparation \& distribution from infrastructure layer devices and provide a unified control plane to deal with global environmental interference for supporting optimal entanglement routing.

\subsection{Three-layer Hierarchical Quantum Internet Architecture}
\label{three_layer_hierarchical_arch}

\begin{figure}[tp]
\centering
\includegraphics[width=3in]{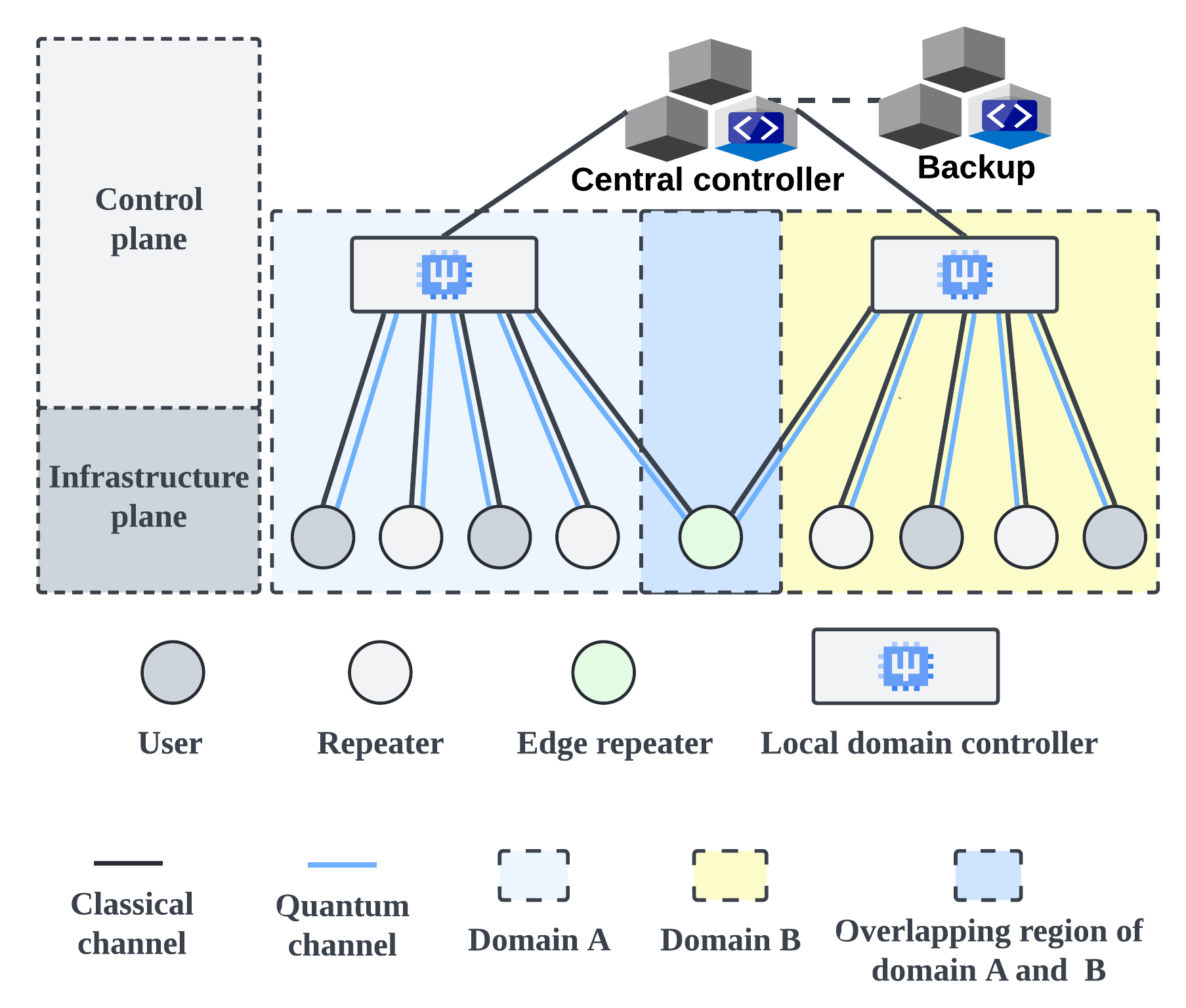}
\caption{Three-layer hierarchical quantum Internet architecture. The hierarchical architecture introduces the central controller and local domain controllers to the quantum Internet. The local domain controller is responsible for centralized entanglement preparation \& distribution. The central controller is responsible for network control and information collection. The edge repeater (green circle) is a medium for domain interaction.}
\label{hierarchical_arch}
\end{figure}

Fig. \ref{hierarchical_arch} shows the three-layer hierarchical quantum Internet architecture. At the top layer, the central controller is connected to local domain controllers through classical channels. At the middle layer, local domain controllers are intermediaries between the top and bottom layers. At the bottom layer, quantum devices are connected to the local domain controller through classical and quantum channels. The central controller and local domain controllers constitute the control plane. The bottom layer, called the infrastructure plane, performs the specific quantum operations.

The central controller is responsible for global network information collection and error control. As for information collection, we design the Central State Matrix (CSM) to collect global information, including device state, quantum memory state, entanglement preparation \& distribution success rate, and entanglement swapping success rate (see details in Appendix \ref{example_LSM_CSM}). With the help of CSM, the central controller can support optimal entanglement routing, devices discovered, and strategy decisions (see Communication Model in Sec. \ref{model_and_protocols}). As for error control, we set entanglement distribution timer $t_d$ and entanglement swapping \& teleportation timer $t_{st}$ to perform time synchronization of quantum operations. The central controller can detect the timeout and prevent the error from damaging the target qubit, thereby preserving the opportunity to retry the communication. We must clarify that the central controller is responsible for one Q-LAN (as shown in Fig. \ref{hierarchical_arch}), and the quantum Internet could include many central controllers that can collaborate.

The local domain controller is responsible for local domain information collection and entanglement preparation \& distribution. As for local domain information collection, we design the Local State Matrix (LSM) to collect local domain information (see details in Appendix \ref{example_LSM_CSM}). Every local domain controller reports the LSM to CSM through the classical channel. As for entanglement preparation \& distribution, the local domain controller performs CEPD so that any two devices in the same domain can be directly distributed entanglement through the quantum channel.

Our control plane design follows the basic requirements of the RFC 9340\cite{kozlowski2023rfc}; as mentioned in the above two paragraphs, the control plane is composed of the central controller and local domain controllers which can collect network information, manage resources, and control the network.

The quantum repeater and user constitute the infrastructure plane. Generally, each quantum device is connected to its local domain controller with classical and quantum channels. However, there is a particular class of quantum devices called edge repeater, located in the overlapping region of two domains and connected to multiple adjacent local domain controllers. The edge repeater is a medium for interaction between domains and performs entanglement swapping to support inter-domain communication\footnote{The inter-domain communication requires swapping to generate long-distance entanglement. The intra-domain communication requires the local domain controller to distribute entanglement for users directly.}. 

Moreover, integrating the existing quantum repeater technology into the hierarchical architecture is an important future work: quantum repeaters are divided into three generations according to the methods used to correct loss and operation errors\cite{muralidharan2016optimal}. We believe that these different types of repeaters can be integrated into the hierarchical architecture. For example, the controller can assist the repeaters in controlling the Quantum Error Correction (QEC).

\subsection{Scalability and Robustness of Hierarchical Architecture}
\label{scalability_robustness}

\subsubsection{Scalability}

The hierarchical quantum Internet architecture is designed for large-scale global quantum Internet and has high scalability. Taking the cellular topology as an example, we discuss the scalability from four perspectives: maintenance cost growth, communication cost growth of the control plane, plug-and-play potential, and automated configuration potential.

\begin{figure}[tp]
\centering
\begin{minipage}[t]{0.49\linewidth}
\centering
\includegraphics[width=1.7in]{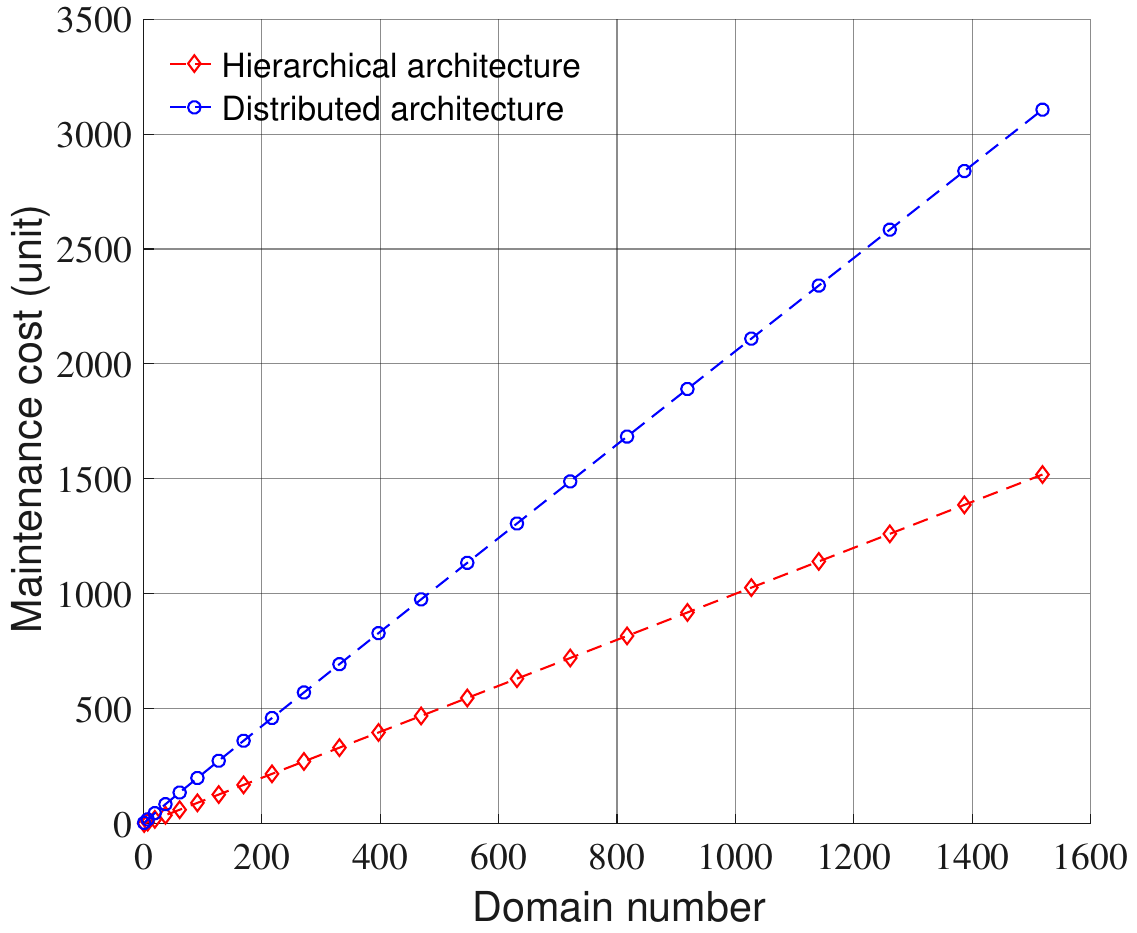}
\caption{Maintenance cost of distributed and hierarchical architecture}
\label{maintenance_cost}
\end{minipage}
\begin{minipage}[t]{0.49\linewidth}
\centering
\includegraphics[width=1.7in]{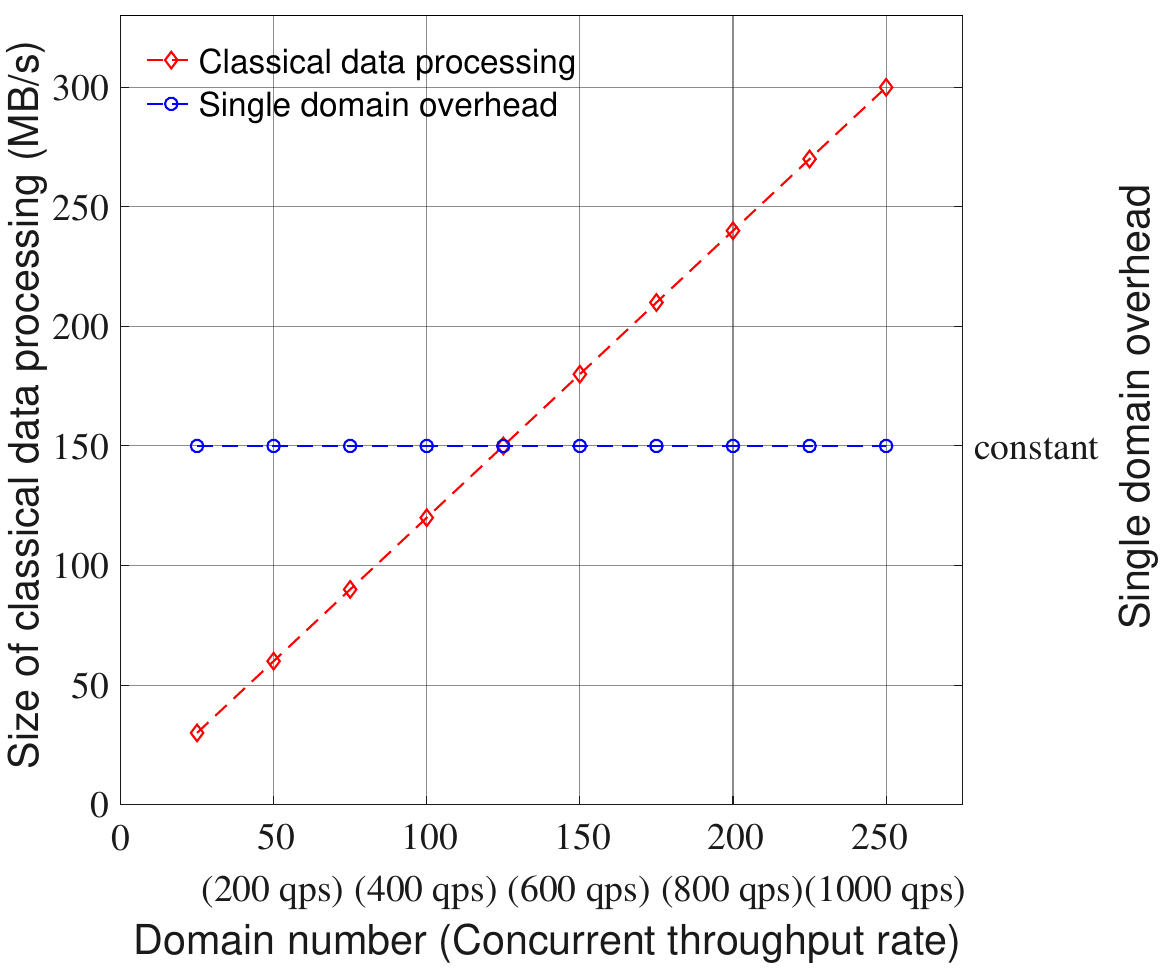}
\caption{Data processing and domain overhead of hierarchical architecture}
\label{data_processing_domain_overhead}
\end{minipage}
\end{figure}

As for maintenance cost growth, we assume that the maintenance cost of one entanglement preparator is 1 unit. In a distributed cellular topology (Fig. \ref{distributed_arch}), a domain has four repeaters as preparators, so the maintenance cost equals the number of repeaters. In a hierarchical cellular topology (Fig. \ref{example_cellular}), a domain has only one local domain controller as a preparator, so the maintenance cost equals the number of local domain controllers. Fig. \ref{maintenance_cost}  shows the increase in maintenance costs with increasing network scale. As can be seen from Fig. \ref{maintenance_cost}, the maintenance cost of the hierarchical architecture is lower than that of the distributed architecture when the scale is the same, and hierarchical architecture has a lower growth rate of maintenance costs as the network expands. Thus, we conclude that the hierarchical architecture is more scalable regarding maintenance cost than the distributed architecture.

As for communication cost growth of the control plane, the data processing capacity of the central controller and the channel bandwidth are important scalability considerations. In the experiment, we calculated that the size of classical data transmission between the control and infrastructure layers in a single quantum communication is about 300 KB (including communication request, BSM result, and control information transmission). Fig. \ref{data_processing_domain_overhead} shows that with the increase of network scale and concurrent communications, the size of classical data processing increases linearly. When the concurrent throughput reaches 1000 qps (qubits per second), the required information processing capability is only 300 MB/s. Moreover, a hierarchical architecture can be easily scaled by increasing the number of domains. Each domain is relatively independent, meaning network expansion does not burden the domain itself. As shown in Fig. \ref{data_processing_domain_overhead}, when the number of users in a domain is fixed, the single domain overhead will not increase, no matter how the network grows. Therefore, from the perspective of communication cost, we can conclude that the scale expansion of hierarchical architecture does not impose an additional burden on the existing infrastructure layer devices, and the total communication overhead shows an acceptable linear growth.

As for plug-and-play potential, in the hierarchical architecture, infrastructure layer devices are only responsible for performing specific quantum operations, while the central controller handles the configuration of devices joining or leaving. The central controller utilizes CSM to collect global network information, such as device name, status, and network topology. This mechanism makes it possible for hierarchical architecture to enable Plug-and-Play as the network expands; for example,  when joining a repeater or user, we connect the new device to the corresponding local domain controller, and then the central controller updates the CSM to allow the new device to be added to the network without requiring a massive update of the infrastructure layer devices.

As for automated configuration potential, because the joining of new devices during network expansion only requires the configuration update on the control plane, the process has the potential to be automated. For example, when the system detects a new device connecting, it can automatically notify the CSM of the central controller for configuration updates, just as new hardware connects to a personal computer.

\subsubsection{Robustness}

The central controller needs to be robust, like the single point of failure problem in classical networks, which can be solved by setting up backup devices. We note that, as shown in Fig. \ref{hierarchical_arch}, since a central controller can manage a Q-LAN composed of many domains, the number of Q-LANs on the Internet equals the number of backup devices required. We also consider the utilization of backup devices; it is possible to make the master central controller and the slave central controller backup for each other in a Q-LAN; the two central controllers can process information simultaneously under normal working conditions. Assuming that the maximum resource usage of a single central controller is 50\% in normal working conditions, if one fails, traffic can be transferred to the remaining controller until the fault is repaired. As for the robustness of links and other devices, the fault of a link or device can only affect a domain or a single device. In addition, the central controller can detect the fault in time, and the status is updated to ``maintain'' in CSM so that communication can avoid faulty devices or links.

\section{Communication Model}
\label{model_and_protocols}

\begin{figure}[tp]
\centering
\includegraphics[width=3in]{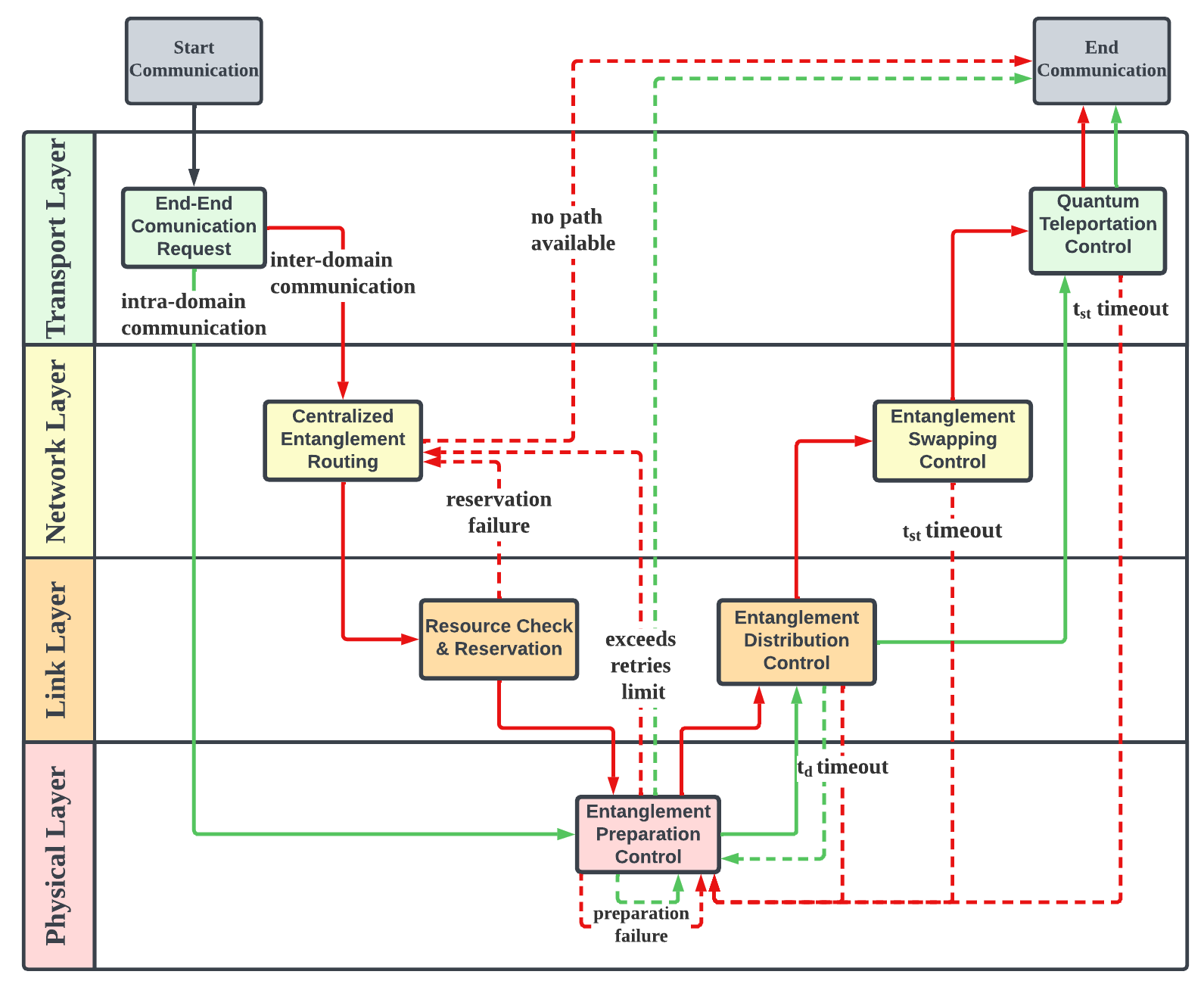}
\caption{The communication model. The solid lines represent the normal communication process, and the dotted lines represent the error control. The green lines are the intra-domain communication, and the red lines are the inter-domain communication.}
\label{communication_modle}
\end{figure}

We propose the communication model of the hierarchical quantum Internet, as shown in Fig. \ref{communication_modle}. All pseudocode of the communication model is shown in\cite{hierarchicalquantumarch2023}. The model uses the same protocol stack as the existing research\cite{li2021building, kozlowski2019towards, dahlberg2019link}, but in terms of protocol content, we consider the hierarchical architecture proposed in this paper. The model starts with the end-end communication request. For intra-domain communication, we only need entanglement preparation, distribution, and quantum teleportation. For inter-domain communication, entanglement routing and swapping are required. In the model, we describe the mechanism of error control. For example, when quantum operations are timeout, we terminate subsequent operations to prevent the target qubit from being destroyed and retry entanglement preparation. When the number of retries exceeds the limit, we choose an alternative path. Error control can avoid the occasionality of node environment interference fluctuations, thus increasing the communication success rate. We should discuss the timer $t_d$ and $t_{st}$ in Fig. \ref{communication_modle}, which are for checking quantum operations timeout. $t_d$  is the entanglement distribution timer that acts on the entanglement distribution process. The timeout of $t_d$  will be triggered when the channel or preparation fails, thereby triggering the local domain controller to retry preparation and distribution. $t_{st}$ is the entanglement swapping \& teleportation timer that acts after entanglement has been distributed to infrastructure layer devices. The timeout of $t_{st}$ will be triggered if the entanglement swapping or teleportation takes too much time (meaning that qubits have most likely decohered), thereby triggering the local domain controller to retry preparation and distribution.

\subsection{Physical Layer}
\label{physical_layer}

\subsubsection{W-state Based Centralized Entanglement Preparation \& Distribution}
\label{W_state_based_CEPD}

\begin{figure*}[tp]
\centering
\includegraphics[width=6in]{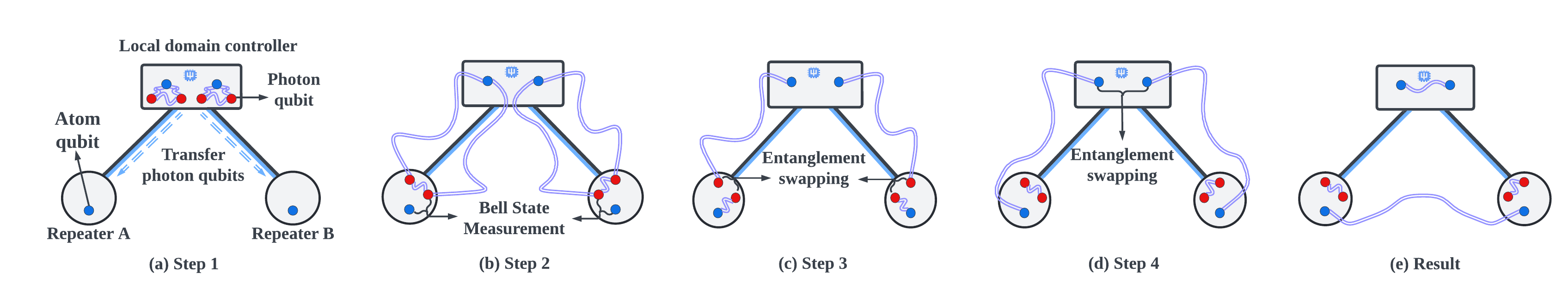}
\caption{W-state Based Centralized Entanglement Preparation \& Distribution. (a) The local domain controller prepares W-state entanglement and transfers photons. (b) Repeaters perform BSM between atom and photon. (c) Repeaters perform entanglement swapping between photons. (d) The local domain controller performs entanglement swapping between atoms. (e) Result of entanglement distribution.}
\label{W_state_CEPD}
\end{figure*}

It is more rational to store quantum information in particles (such as $^{171}Yb$,$^{13}C$, and $^{87}R$) for long-distance quantum communication because they can provide longer qubit storage time\cite{alden2010calculation, abobeih2018one, reiserer2014quantum}. However, the question is how we can distribute atom-atom entangled pairs without requiring repeaters to perform entanglement preparation. We propose W-state Based Centralized Entanglement Preparation \& Distribution. The main idea of W-state Based CEPD is that the local domain controller utilizes tripartite W-state entanglement characteristics\cite{kiesel2003three} (when one qubit is destroyed, the other two qubits remain entangled) to distribute atom-atom entangled pairs without requiring repeaters to perform entanglement preparation. Fig. \ref{W_state_CEPD} shows the W-state Based CEPD. There are four steps to distribute atom-atom entangled pairs between repeaters A and B:

{\bfseries Step 1:} In Fig. \ref{W_state_CEPD}(a), the local controller prepares two groups of atom-photon-photon W-state entanglement and distributes photons to repeaters. The feasibility of preparing atom-photon-photon W-state entanglement has been verified in\cite{alden2010calculation, alden2012yb}.

{\bfseries Step 2:} In Fig. \ref{W_state_CEPD}(b), repeaters perform BSM between photon and local atom qubit. We gain four groups of atom-photon entanglements after BSM.

{\bfseries Step 3:} In Fig. \ref{W_state_CEPD}(c), repeaters perform entanglement swapping between photons. We gain two groups of atom-atom entanglements between the local domain controller and repeaters after entanglement swapping.

{\bfseries Step 4:} In Fig. \ref{W_state_CEPD}(d), the local controller performs entanglement swapping between two local atom qubits. Finally, as shown in Fig. \ref{W_state_CEPD}(e), we obtain the atom-atom entangled pair between repeaters.

For converting the W-state to two EPR pairs in step 2, we sketch the derivation to calculate the probability of success for one of the W-states - the other is similar. The W-state is denoted as $\ket{W}_{q_{p_1}q_{p_2}q_{a_{lc}}} = \frac{1}{\sqrt{3}}(\ket{001}+\ket{010}+\ket{100})$ , and the atom qubit in the repeater is denoted as $\ket{\psi}_{q_{a_r}}$. We perform quantum operations $C_{NOT}(control:H(\ket{\psi}_{q_{a_r}}), target:\ket{q_{p_1}})$; the process can be expressed as formula \ref{w_state_derive_1}-\ref{w_state_derive_result_2}.
\begin{equation}
\begin{aligned}
&\ket{\psi}_{q_{a_r}} \otimes \ket{W}_{q_{p_1}q_{p_2}q_{a_{lc}}} \qquad \qquad \qquad \qquad \qquad \qquad\\
= & \ket{\psi}_{q_{a_r}} \otimes \frac{1}{\sqrt{3}}(\ket{0}_{q_{p_1}} \otimes \ket{01}_{q_{p_2}q_{a_{lc}}} + \ket{0}_{q_{p_1}} \otimes \ket{10}_{q_{p_2}q_{a_{lc}}} \\ 
\quad & + \ket{1}_{q_{p_1}} \otimes \ket{00}_{q_{p_2}q_{a_{lc}}}) \\
= & \frac{1}{\sqrt{3}}(\ket{\psi}_{q_{a_r}}\ket{0}_{q_{p_1}} \otimes \ket{01}_{q_{p_2}q_{a_{lc}}} + \ket{\psi}_{q_{a_r}}\ket{0}_{q_{p_1}} \otimes \ket{10}_{q_{p_2}q_{a_{lc}}} \\
\quad & + \ket{\psi}_{q_{a_r}}\ket{1}_{q_{p_1}} \otimes \ket{00}_{q_{p_2}q_{a_{lc}}})
\label{w_state_derive_1}
\end{aligned}
\end{equation}
Assuming initially $\ket{\psi}_{q_{a_r}} = \ket{0}_{q_{a_r}}$, according to formula \ref{w_state_derive_1} and the quantum operations, after applying $H$ and the $C_{NOT}$ (basically operations in a typical BSM), the pair of qubits $q_{a_r}$ and $q_{p_1}$ will be in a Bell State, where we have a two-thirds probability of getting the result of formula \ref{w_state_derive_result_1} (two EPR pairs) and a one-third probability of getting the result of formula \ref{w_state_derive_result_2}.
\begin{equation}
\begin{aligned}
\ket{\Phi^+}_{q_{a_r}q_{p_1}} \otimes \ket{\Psi^+}_{q_{p_2}q_{a_{lc}}}
\label{w_state_derive_result_1}
\end{aligned}
\end{equation}
\begin{equation}
\begin{aligned}
\ket{\Psi^+}_{q_{a_r}q_{p_1}} \otimes \ket{00}_{q_{p_2}q_{a_{lc}}}
\label{w_state_derive_result_2}
\end{aligned}
\end{equation}

For the entanglement swapping, we take step 3 as an example (step 4 is similar). According to the previous derivation, the result of the successful execution of step 2 can be written as $\ket{\Phi^+}_{q_{a_r}q_{p_1}}  \otimes \ket{\Psi^+}_{q_{p_2}q_{a_{lc}}}$. We utilize the result of step 2 to perform entanglement swapping. The entanglement swapping projects $q_{a_{lc}}$ and $q_{a_r}$ also onto an EPR pair.

Fig. \ref{W_state_CEPD} shows the qualitative evolution of W-state Based CEPD. Considering the environmental interference, we give the probability derivation (due to device/channel-level factors) of evolution from W-state to Bell State in W-state Based CEPD, as described below. We assume that the success rate of W-state preparation is $p_{\ket{W}}$ and the success probability of quantum channel transmission is $p_{qchannel}$.  $p_{\ket{W}}$ mainly depends on the performance of the local controller, and $p_{qchannel}$ depends on the two interference parameters of the quantum channel, loss\_init\_rate and loss\_noise. In step 1, the local controller should prepare two W-state entanglements and transmit photons from two different channels, so its success probability $p_1$ can be expressed as formula \ref{step_1_p}. We assume the success rate of BSM is $p_{bsm}$, the success rate of photons' entanglement swapping is $p_{p\_swap}$, and the success rate of atoms' entanglement swapping is $p_{a\_swap}$, which are mainly affected by the dephasing rate. In step 2, two repeaters, A and B, should perform BSM between atom and photon, so its success probability $p_2$ can be expressed as formula \ref{step_2_p}. In step 3, two repeaters, A and B, should perform entanglement swapping between photons, so its success probability $p_3$ can be expressed as formula \ref{step_3_p}. In step 4, the local controller should perform entanglement swapping between atoms, so its success probability $p_4$ can be expressed as formula \ref{step_4_p}. After completing four steps, we obtain the Bell State (EPR pairs), so the probability $p_{\ket{EPR}}$ can be expressed as formula \ref{step_result_p}. Moreover, in the simulation experiment, we also fully considered the impact of environmental interference on W-state Based CEPD. See Appendix \ref{simulation_W_state} for the derivation of experimental parameters.

\begin{equation}
\begin{aligned}
p_{1} = \prod_{i=1}^{2}p_{\ket{W_i}} \times \prod_{i=1}^{2}p_{qchannel_i}(loss\_init,  loss\_noise)
\label{step_1_p}
\end{aligned}
\end{equation}

\begin{equation}
\begin{aligned}
p_{2} = &p_{bsm}(\ket{\psi_{R_A\_atom}},\ket{\psi_{R_A\_photon1}},dephasing) \times \\ 
&p_{bsm}(\ket{\psi_{R_B\_atom}},\ket{\psi_{R_B\_photon1}},dephasing)
\label{step_2_p}
\end{aligned}
\end{equation}

\begin{equation}
\begin{aligned}
p_{3} = &p_{p\_swap}(\ket{\psi_{R_A\_photon1}},\ket{\psi_{R_A\_photon2}},dephasing) \times \\ 
&p_{p\_swap}(\ket{\psi_{R_B\_photon1}},\ket{\psi_{R_B\_photon2}},dephasing)
\label{step_3_p}
\end{aligned}
\end{equation}

\begin{equation}
\begin{aligned}
p_{4} = p_{a\_swap}(\ket{\psi_{LC\_atom1}},\ket{\psi_{LC\_atom2}},dephasing)
\label{step_4_p}
\end{aligned}
\end{equation}

\begin{equation}
\begin{aligned}
p_{\ket{EPR}} = \prod_{i=1}^{4}p_{i}
\label{step_result_p}
\end{aligned}
\end{equation}

We note that W-state Based CEPD requires extra quantum operations and photon transmissions (Table \ref{distribution_schemes}), which may increase the probability of quantum decoherence. In order to analyze the impact of the adverse factor, we propose the Double-photon Based Centralized Entanglement Preparation \& Distribution (Double-photon Based CEPD) (see Appendix \ref{double_photon_CEPD}) as a comparison of W-state Based CEPD. The Double-photon Based CEPD adopts the simplest double-photon entanglement distribution in the hierarchical architecture without extra quantum operations and photon transmissions. The experiment in Sec. \ref{performance_of_entanglement_pd} shows that W-state Based CEPD is superior to Double-photon Based CEPD and Double-photon Based DEPD when the efficiency of atomic memory is sufficient. Also, with the increase in the efficiency of atomic memory, the performance of W-state Based CEPD will be better and better. In the real world, technology determines the efficiency of atomic memory\cite{zhao2009long}. With the development of technology, the advantages of W-state Based CEPD will be more obvious.

\subsubsection{Entanglement preparation control}
\label {entanglement_preparation_control}
The end-end path is required by entanglement preparation. We give an example as shown below:\\
\centerline{{\footnotesize $U_A[LC_A(cm_1,cm_2),um_1] \rightarrow$ }}
\centerline{{\footnotesize $R_A[LC_A(cm_1,cm_2),LC_B(cm_1,cm_2),rm_1,rm_2]  \rightarrow$}}
\centerline{{\footnotesize $U_B[LC_B(cm_1,cm_2),um_1]$}}
U represents the user, R represents the repeater, LC represents the local domain controller, and cm/rm/um represents the memory of the local domain controller/repeater/user. The example shows a quantum communication path from $U_A$ to $U_B$. $LC_A$ utilized $cm_1$ and $cm_2$ to prepare W-state entanglement for entanglement distribution between $U_A$'s $um_1$ and $R_A$'s $rm_1$. $LC_B$ utilized $cm_1$ and $cm_2$ to prepare W-state entanglement for entanglement distribution between $R_A$'s $rm_2$ and $U_B$'s $um_1$. And then, $R_A$ performs entanglement swapping between $rm_1$ and $rm_2$ to generate the long-distance entanglement for quantum teleportation.

The central controller informs local domain controllers which belong to the path to perform entanglement preparation. If the entanglement preparation succeeds, the CSM and LSM update PreparationState and other relevant information, and local domain controllers start the entanglement distribution process. If the entanglement preparation fails, the CSM and LSM update LinkState (representing the success rate of entanglement preparation \& distribution; see formal definition in Appendix \ref{example_LSM_CSM}), and local domain controllers retry entanglement preparation. When the number of retries exceeds the limit, the CSM and LSM set the DeviceState (representing the device as normal or faulty; see formal definition in Appendix \ref{example_LSM_CSM}) of relevant local domain controllers to ``maintain''. For intra-domain communication, the communication fails. For inter-domain communication, the central controller restarts entanglement routing to select an alternative path.

\subsection{Link Layer}
\label{link_layer}
\subsubsection{Resource check \& reservation}
The central control confirms the device state and finds idle memories according to the CSM and $Path_{middle}$. $Path_{middle}$ is a middle-state path given by the entanglement routing algorithm without memories of local domain controllers and repeaters (see $Path_{middle}$ example in Sec. \ref{CER}). Devices of the $Path_{middle}$ reserve memories resource according to instructions from the central controller. The CSM and LSM set the relevant MemoryState to ``occupy''. If the reservation succeeds, we receive a completed communication path (see the example in Sec. \ref{entanglement_preparation_control}). If the reservation fails, the central controller restarts entanglement routing to select an alternative path.

\subsubsection{Entanglement distribution control}
After resource reservation and entanglement preparation are completed, local domain controllers begin to distribute entanglement to repeaters according to the process of W-state Based CEPD (as shown in Fig. \ref{W_state_CEPD}). Considering the possibility of timeouts due to poor channel quality (meaning possible decoherence of qubits), so we set an entanglement distribution timer $t_d$ to perform the time synchronization of entanglement distribution. In the distribution process, the central controller performs the entanglement distribution control, which will confirm the distribution results of repeaters and adopt different processing methods according to the distribution results. There are two kinds of results. The first case is as follows: if all entanglement distributions succeed and the $t_d$ is not timeout, the central controller informs the source user to perform quantum teleportation for intra-domain communication or informs repeaters to perform entanglement swapping for interdomain communication. Meanwhile, CSM and LSM update DistributionState and other relevant information. The second case is as follows: if failure or timeout happens during the entanglement distribution, the central controller terminates subsequent operations and retries entanglement preparation in the current path to prevent the target qubit from being broken. When the number of retries exceeds the limit, CSM and LSM set the DeviceState of failed devices to ``maintain". Then, the central controller restarts entanglement routing to select an alternative path for inter-domain communication or announces that communication fails for intra-domain communication. We should mention that during the distribution process, the central controller will record whether the entanglement distribution of repeaters succeeded or failed. Then, CSM and LSM can update the LinkState according to the result of entanglement distribution. In this way, LinkState can reflect the success rate of the repeater's entanglement distribution, which will be used as one of the basis for future routing.

\subsection{Network Layer}
\label{network_layer}
\subsubsection{Centralized entanglement routing}
\label{CER}

\begin{figure}[tp]
\centering
\includegraphics[width=3in]{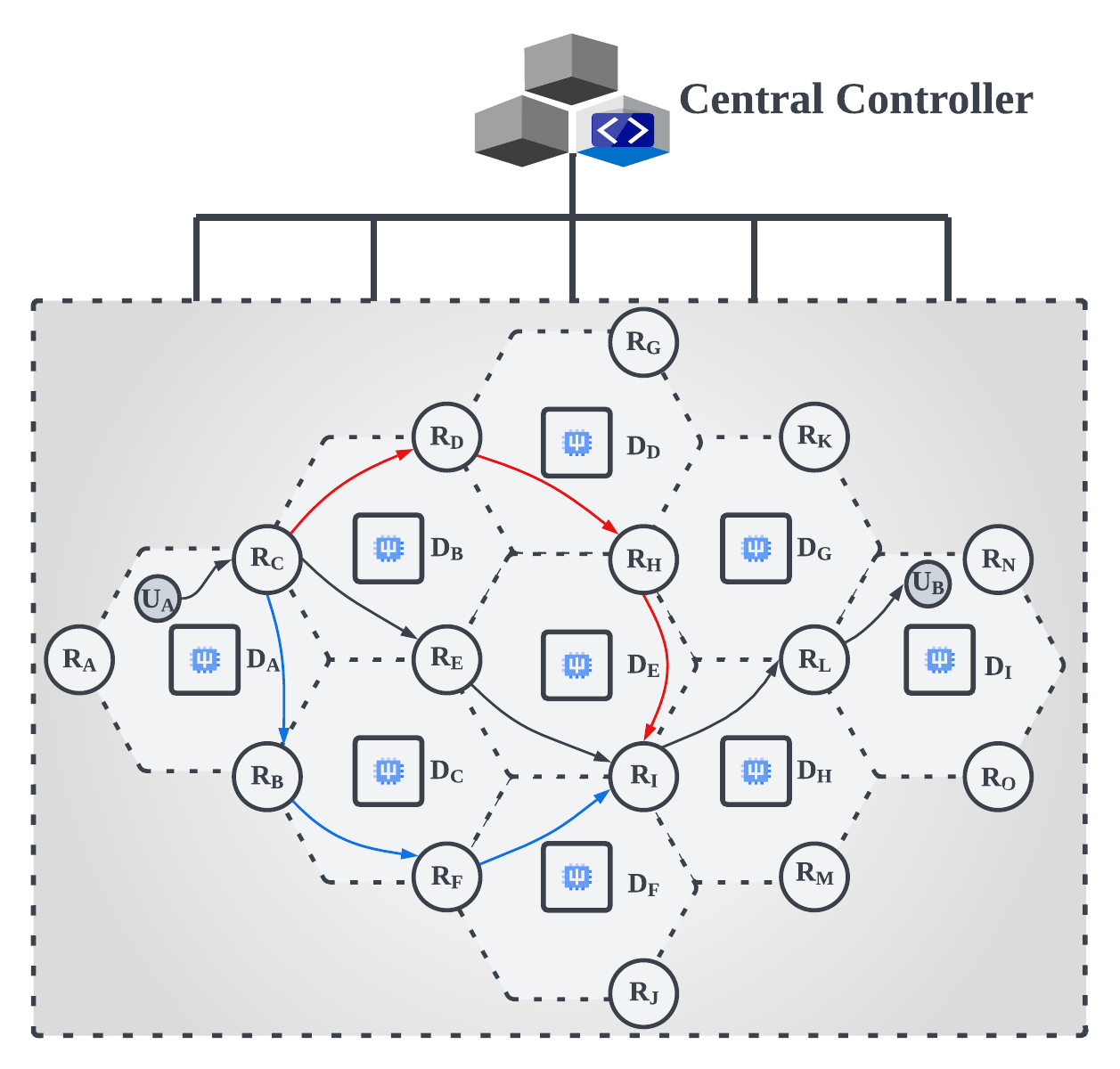}
\caption{Example of hierarchical cellular topology. ``D'' is Domain, ``U'' is User, and ``R'' is Repeater. The central controller connects to each local domain controller through classical channels. The figure omits the identification of classical and quantum channels between the local domain controller and quantum devices.}
\label{example_cellular}
\end{figure}

CER utilizes hops, entanglement preparation \& distribution success rate, and entanglement swapping success rate to evaluate the paths. The path with the max score is the optimal path for quantum communication. The core process of CER is expanding hops of all shortest paths through iteration rules and selecting the optimal path according to the score. As shown in Alg. \ref{algo_CER}, CER is divided into three phases: 

{\bfseries Generating candidate paths:} We bring Domain Shortest Path Table (DSPT) and Domain Edge Repeater Table (DERT) into the central controller for recording the shortest path between domains and information of edge repeaters. DSPT and DERT are implemented by Struct (see source code in\cite{hierarchicalquantumarch2023}). With the help of DSPT and DERT, we can directly obtain the shortest paths set of the first iteration. For example, the communication between $U_A$ and $U_B$ in Fig. \ref{example_cellular} has six shortest paths. We take one of them as an example:\\
\centerline{{\footnotesize $U_A[LC_A,um_1] \rightarrow R_C[LC_A,LC_B] \rightarrow R_E[LC_B,LC_E] \rightarrow$ }}
\centerline{{\footnotesize $R_I[LC_E,LC_H] \rightarrow R_L[LC_H,LC_I] \rightarrow U_B[LC_I,um_1]$}}
This example is the black path in Fig. \ref{example_cellular}. It represents that the $LC_A$ performs entanglement distribution for $U_A$ and $R_C$,  $LC_B$ performs entanglement distribution for $R_C$ and $R_E$, and the rest can be done similarly. Besides, all paths given by CER are middle-state paths, which only contain the memories of users. The communication request process gives memories of users initially, while the resource check \& reservation process gives memories of local domain controllers and repeaters after the entanglement routing. CER adds one hop for each path per iteration to expand the set of candidate paths. In each iteration, CER replaces each repeater of the previous round paths. The replacement rule is that two new repeaters are selected from the neighboring nodes around the old repeater to replace the old repeater, thus forming a new path. The new path has the opportunity to avoid bad links and nodes of the old path. For example, in Fig. \ref{example_cellular}, $R_E[LC_B,LC_E]$ of the black path is replaced to generate red and blue paths.

 \IncMargin{1em}
\begin{algorithm}\SetKwInOut{Input}{input}\SetKwInOut{Output}{output}\SetKw{Break}{break}
\let\oldnl\nl 
{\fontsize{10pt}{15pt}\selectfont
\newcommand{\nonl}{\renewcommand{\nl}{\let\nl\oldnl}}	
	\Input{$U_{src}$, $U_{dst}$} 
	\Output{optimal $Path_{middle}$}
	 \tcp{$T_{dsp}$: domain shortest path table}
	 \tcp{$T_{der}$: domain edge repeater table}
	 \tcp{N: recursion number}
	 \tcp{$Path_{pr}$: paths of previous round}
	 \tcp{$Path_{cr}$: paths of current round}
	 $D_{src}$, $D_{dst}$ = FindDomain($T_{csm}$, $U_{src}$, $U_{dst}$)\;     
	 $Path_{array}$ = FindPaths($T_{dsp}$, $T_{der}$, $D_{src}$, $D_{dst}$)\;
	 $Path_{pr}$ $\leftarrow$ $Path_{array}$\;
	 \While{N $>$ 0}{
	      \For{path $\in$ $Path_{pr}$}{
	          \For{repeater $\in$ path}{
	              nodes = FindReplaceNodes($T_{csm}$)\;
	              \If{nodes is not none}{
	                  $Path_{cr}$ $\leftarrow$ AddNewPath\;
	              }
	          } 
	      }
	      \If{$Path_{cr}$ is none}{
	          \Break
	      }
	      $Path_{array}$ $\leftarrow$ $Path_{cr}$\;
	      Clear($Path_{pr}$)\;
	      $Path_{pr}$ $\leftarrow$ $Path_{cr}$\;
	      N = N - 1\;
	  }
	  EliminateInvalidPath($Path_{array}$)\;
	  \For{path $\in$ $Path_{array}$}{
	      \For{repeater $\in$ path}{
	          SwappingSuccessRate = Result({$T_{csm}$})\;
	          LinkState = Result({$T_{csm}$})\;
	          $Score_R$ = SwappingSuccessRate $\times$  0.3 + \\
	          \nonl \qquad \qquad \quad LinkState $\times$ 0.7\;
	          $Score_P$ = $Score_P$ + $Score_R$\;
	      }        
	      $Score_P$ = $\frac{Score_P}{hops}$\;
	      $Score_{array}$ $\leftarrow$ $Score_P$\;
	  }
	  sortpaths = Sort($Score_{array}$, $Path_{array}$, hops)\;
	\While{true}{
	    $Path_{middle}$ = sortpaths.pop(0)\;
	    Path = ResourceCheck\&Reserve($Path_{middle}$)\;
	    \If{Path != null}{
	        start entanglement preparation process\;
	        \Break
	   }
	   \ElseIf{sortpaths == null}{
	       announce no path found\;
	       announce communication failed\; 
	       \Break
	   }
	 }
\caption{Centralized Entanglement Routing.}
\label{algo_CER} 
}
\end{algorithm}
\DecMargin{1em} 

{\bfseries Eliminating invalid paths:} There might be invalid paths in the set of candidate paths. For example, the blue path in Fig. \ref{example_cellular} is invalid because $U_A$, $R_C$, and $R_B$ are three consecutive devices in the same domain. The combination of $U_A$, $R_C$, and $R_B$ is equivalent to the combination of $U_A$ and $R_B$ because even adding $R_C$ as an intermediate node cannot bypass the faulty channels of $U_A$-$LC_A$ and $R_B$-$LC_A$. However, we should note that a path with three discontinuous repeaters in the same domain is valid. For example, {\footnotesize $R_E[LC_C,LC_B] \rightarrow R_D[LC_B,LC_D] \rightarrow R_H[LC_D,LC_E] \rightarrow R_I[LC_E,LC_H]$} in Fig. \ref{example_cellular}, $R_E$, $R_H$, and $R_I$ are three discontinuous repeaters in the same domain, and this path seems to be equivalent to $LC_E$ directly perform entanglement distribution for $R_E$ and $R_I$. However, when the channel between $R_E$ and $LC_E$ is faulty, this extended path can bypass the faulty channel.

{\bfseries Evaluating:} CER obtains SwappingSuccessRate (representing entanglement swapping success rate) and LinkState (representing entanglement preparation \& distribution success rate) from CSM. SwappingSuccessRate is mainly determined by the dephasing rate, while the LinkState is mainly determined by quantum channel quality if the entanglement source is stable. According to our experiment, the influence on the communication of quantum channel quality is about twice than dephasing rate (see Appendix \ref{channel_quality_and_dephasing}). Therefore, we set the SwappingSuccessRate weight to 0.3 and LinkState weight to 0.7 in the evaluation. CER evaluates repeaters according to formula \ref{score_repeater}. The path's score is the sum of the repeaters' scores divided by hops, as shown in the formula \ref{score_path}. Finally, CER sorts paths according to scores and hops, the path with the highest ranking is selected for resource reservation. If the resource reservation succeeds, CER starts the entanglement preparing process; otherwise, the subsequent paths are selected for resource reservation until there are no paths in the candidate set, and the communication fails.

\begin{equation}
Score_R = SwappingSuccessRate \times 0.3 + LinkState \times 0.7
\label{score_repeater}
\end{equation}
\begin{equation}
Score_P =\frac{\sum\limits_{i=0}^{n}Score_{R_i}}{hops}
\label{score_path}
\end{equation}

\subsubsection{Entanglement swapping control} 
\begin{figure}[tp]
\centering
\includegraphics[width=3in]{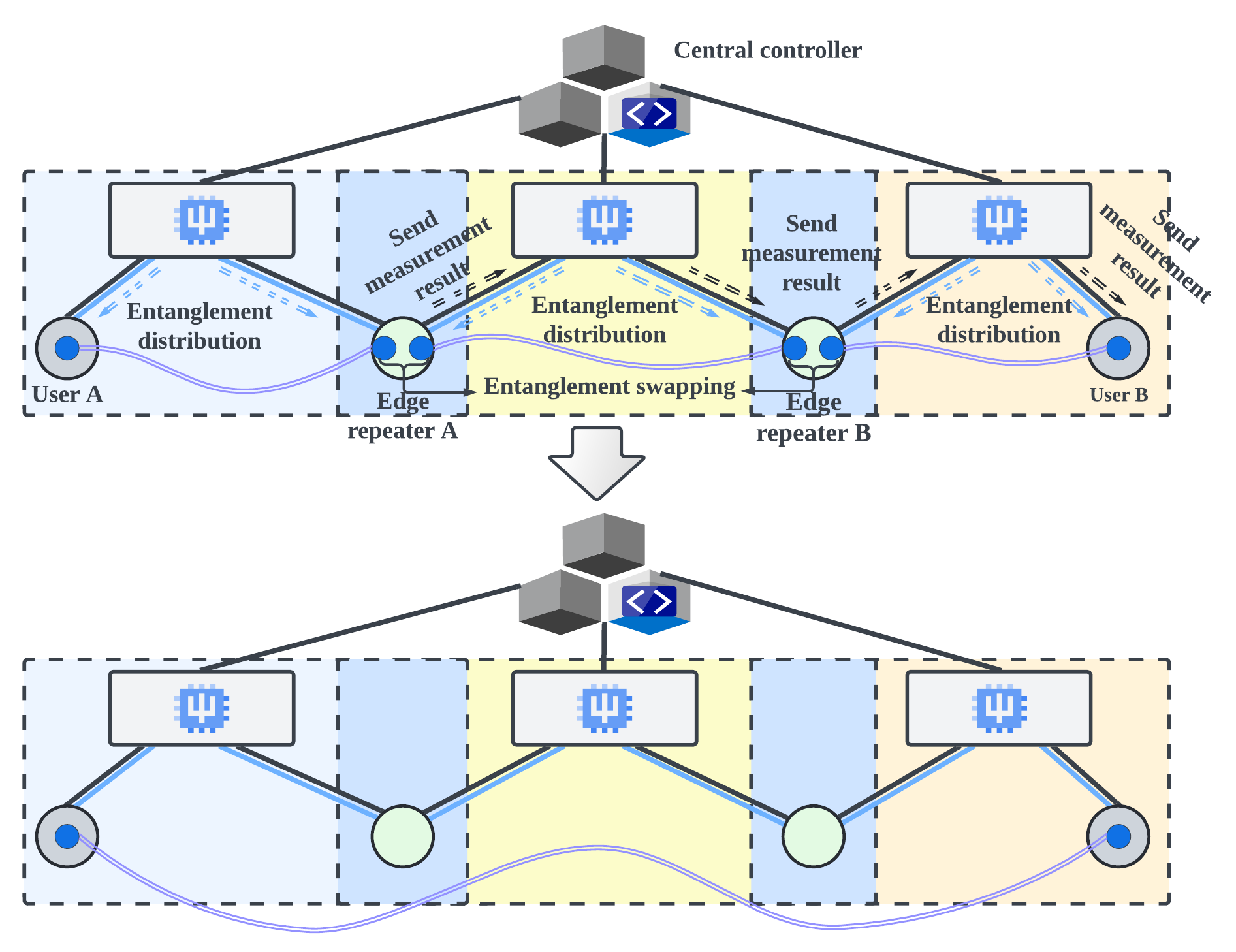}
\caption{Entanglement swapping control. Local domain controllers perform entanglement preparation \& distribution. Edge repeaters perform entanglement swapping.}
\label{inter_domain_swapping}
\end{figure}

After the entanglement distribution, qubits are deposited into memory with a new life cycle, so we set a new timer $t_{st}$ to perform time synchronization of entanglement swapping \& teleportation. The central controller successively informs repeaters to perform entanglement swapping, as shown in Fig. \ref{inter_domain_swapping}. At the same time, CSM and LSM update SwappingSuccessRate according to the entanglement swapping result. If all operations succeed and the $t_{st}$ is not timeout, CSM and LSM update SwappingState, and the central controller informs the source user to start the quantum teleportation process. Otherwise, the central controller terminates subsequent operations and retries entanglement preparation in the current path to prevent the target qubit from being broken. When the number of retries exceeds the limit, CSM and LSM set the DeviceState of failed devices to ``maintain''. And then, the central controller restarts entanglement routing to select an alternative path.

\subsection{Transport Layer}
\label{transport_layer}
 
\subsubsection{End-end communication request} 
Because communication can only start when the source confirms that the destination has idle resources and agrees to receive quantum information, it is necessary to establish the connection between users before communication begins. The communication request and reply carry idle memory information of users. The local domain controller receives the communication request from the source user and identifies the communication type according to LSM. For intra-domain communication, the request is directly transmitted to the destination user. And then, if the reply from the destination user accepts the communication, the local domain controller generates an intra-domain communication path and starts the entanglement preparation and quantum teleportation. For inter-domain communication, the communication request is transmitted to the destination user via the central controller. If the reply of the destination user accepts the communication, the central controller records the idle memories of users and starts the entanglement routing process.

\subsubsection{Quantum teleportation control} 
\begin{figure}[tp]
\centering
\includegraphics[width=3in]{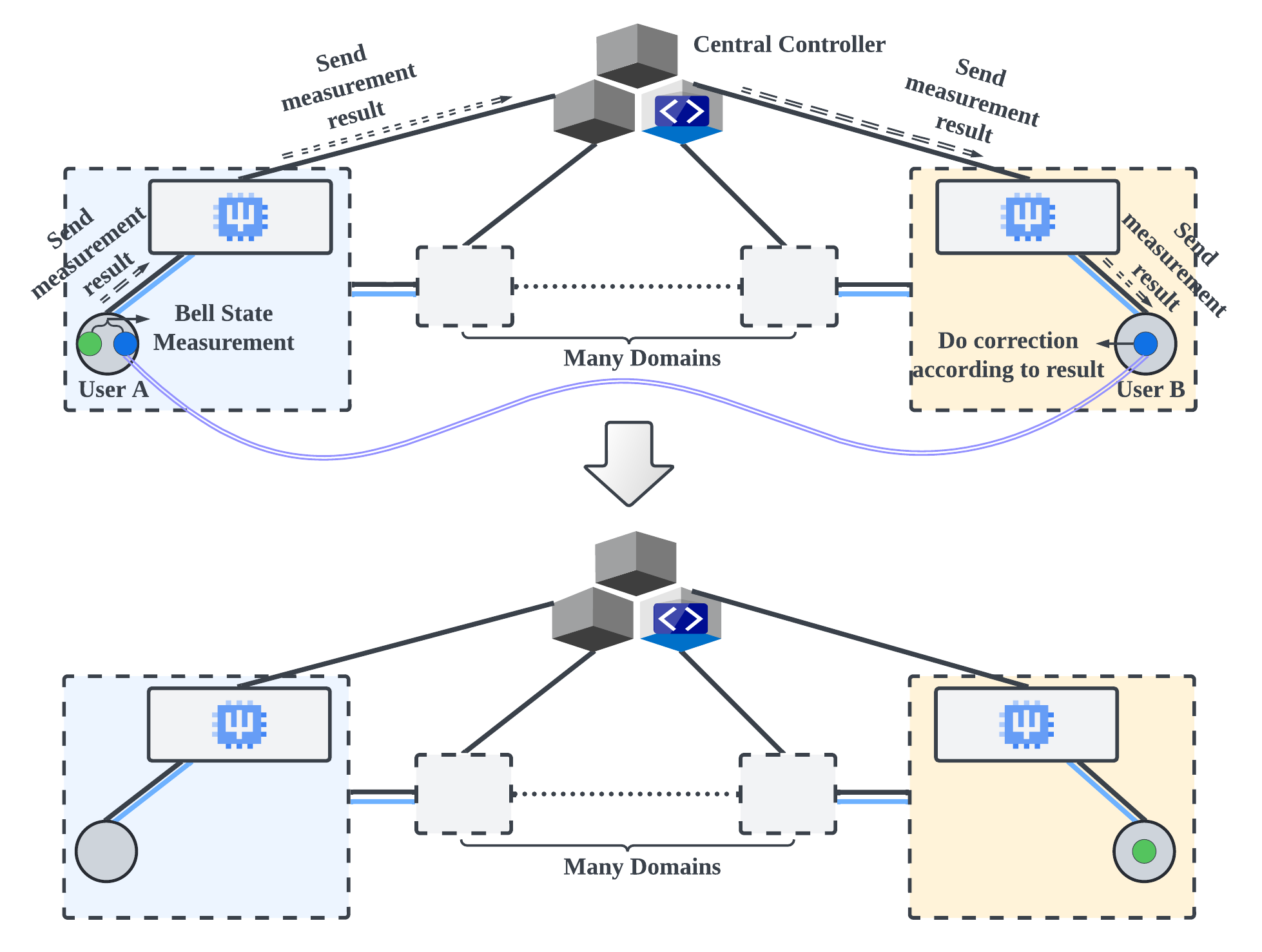}
\caption{Quantum teleportation control. The Source user performs BSM. BSM result is transferred from source to destination via the central controller for qubit correction.}
\label{inter_domain_teleportation}
\end{figure}
The central controller checks the $t_{st}$, which is inherited from entanglement swapping control. On the one hand, if the $t_{st}$ is not timeout, the central controller informs the source user to perform BSM between the target qubit and the one qubit of the entanglement, as shown in Fig. \ref{inter_domain_teleportation}. The source user sends the measurement result to the destination user for qubit correction. If the correction succeeds, the target qubit is successfully transmitted. CSM and LSM update TeleporationState, and the central controller resets all related memories. If the correction fails, the central controller announces that the communication fails because the target qubit has been broken. On the other hand, if the $t_{st}$ is timeout, the central controller retries entanglement preparation in the current path. When the number of retries exceeds the limit, the central controller restarts entanglement routing to select an alternative path.

\section{Evaluation}
\label{evaluation}
\subsection{Implementation}
We need to point out that the evaluation results in this paper are based on simulation. We utilize NetSquid\cite{netsquid2021} to simulate and evaluate the hierarchical quantum Internet architecture. All of our code is published in\cite{hierarchicalquantumarch2023}. Besides, we provide a use case of completed end-end communication in\cite{hierarchicalquantumarch2023}. Readers can adjust the use case parameters to observe the mechanism of hierarchical quantum Internet. NetSquid is a quantum network simulation platform that provides qubits, quantum memory, quantum operation, scalable quantum protocol, and quantum error model. Here, we explain the environmental parameters and evaluation metrics:

{\bfseries Depolarizing rate:} It represents the probability that the qubit in quantum memory depolarizes with time. We can control the memory efficiency by adjusting the depolarizing rate.

{\bfseries Dephasing rate:} It represents the probability that the qubit dephases with time. We can control the entanglement swapping success rate by adjusting the dephasing rate.

{\bfseries Q-channel loss init rate:} It represents the initial probability of losing a photon once it enters a quantum channel. Q-channel loss init rate is a parameter of channel interference.

{\bfseries Q-channel loss noise:} It represents the noise of the quantum channel, and the unit is dB/km. Q-channel loss noise is also a parameter of channel interference. We can control the quantum channel quality by adjusting the Q-channel loss init rate and Q-channel loss noise.

{\bfseries Fidelity:} It represents the fidelity of qubit transmission. We can evaluate quantum communication by the mean and standard error of fidelity.

{\bfseries Throughput:} It represents the throughput of quantum communication, and the unit is qubits per second (qps). We get the throughput by counting the number of successful target qubits transmissions in unit time. The throughput is also an evaluation metric of quantum communication.

\subsection{Performance of Entanglement Preparation \& Distribution}
\label{performance_of_entanglement_pd}

We design an end-end quantum communication experiment to compare the performance of W-state Based CEPD, Double-photon Based CEPD, and Double-photon Based DEPD. Features of three schemes are concluded in Table \ref{distribution_schemes}. Because the entanglement distribution solution is strongly coupled with the architecture, we selected the shortest path (e.g., the black path in Fig. \ref{example_cellular}) from user A to B in Fig. \ref{example_cellular} (for DEPD) and Fig. \ref{distributed_arch} (for CEPD) for the experiment. 

We should explain the simulation parameter settings. Firstly, the single channel length in CEPD and DEPD is 100 km to ensure fairness in photon transmission. Secondly, The distance of users in CEPD and DEPD is the same to ensure the fairness of the communication. Thirdly, CEPD and DEPD use the same geographically shortest path, which means that the devices involved in the paths of different schemes are in the same position. However, because the hierarchical and distributed architecture fundamentally differ in topology, the repeater hops of the shortest path in CEPD is 4, and in DEPD is 9. It should be noted that under the same communication distance, CEPD's fewer repeater hops reflect the advantages of the hierarchical architecture, not a factor affecting fairness. Finally, regarding the setting of environmental interference parameters, as for Double-photon Based CEPD and DEPD, the depolarizing rate is 0.1, the dephasing rate is 0.01, the Q-channel loss init rate is 0.0001, and the Q-channel loss noise is $1.0\times10^{-5}$ dB/km; as for W-state Based CEPD, the environmental interference setting is shown in Appendix \ref{simulation_W_state}.

The result is shown in Fig. \ref{entanglement_pd_solution}. According to the comparison of fidelity, Double-photon Based CEPD is better than Double-photon Based DEPD. Double-photon Based DEPD and Double-photon Based CEPD both takes the optical memory for storing qubit, but CEPD utilizes local domain controllers to perform centralized entanglement preparation \& distribution, which reduces the number of quantum operations during the entanglement distribution, thereby improving the efficiency. As for W-state Based CEPD, when the efficiency of atomic memory is only twice that of optimal memory, W-state Based CEPD is not superior to the other two solutions, which is caused by the negative impact of extra quantum operations and photon transmissions. However, with the increase in the memory efficiency ratio, the fidelity of W-state Based CEPD continues to rise. When the memory efficiency ratio exceeds 5, W-state Based CEPD is much better than the other two schemes. The reason is that when the efficiency of atomic quantum memory is sufficient, the positive impact brought by the long-life qubits outweighs the negative impact.

In conclusion, the efficiency of CEPD is 11.5\% higher than DEPD on average, a minimum of 3\% higher with Double-photon Based CEPD, maximum of 37.3\% higher with W-state Based CEPD. The hierarchical architecture provides a unified entanglement preparator (local domain controller) that enables devices in the same domain to obtain entanglement pairs quickly, thereby making DEPD performs better than CEPD.

\begin{table}
\centering
\caption{Information for schemes of entanglements preparation \& distribution}
\begin{tabular}{|c|c|c|c|}
\hline  
{\bfseries Schemes} & {\bfseries Entanglement} & \makecell[c]{{\bfseries Entanglement} \\ {\bfseries preparator} \\ {\bfseries (Type of} \\ {\bfseries distribution)}} & \makecell[c]{{\bfseries Operations } \\    {\bfseries \& Photon} \\ {\bfseries transmissions} \\ (between $R_C$ \\ and $R_I$ \\ in Fig. \ref{example_cellular} and \\ Fig. \ref{distributed_arch})} \\
\hline 
\makecell[c]{W-state \\ CEPD} & \makecell[c]{atom \\ (long life)} & \makecell[c]{local domain \\ controller \\ (centralized)} & 11 \& 8 \\
\hline
\makecell[c]{Double-photon \\ CEPD} & \makecell[c]{photon \\ (short life)} & \makecell[c]{local domain \\ controller \\ (centralized)} & 1 \& 4 \\
\hline
\makecell[c]{Double-photon \\ DEPD} & \makecell[c]{photon \\ (short life)} & \makecell[c]{quantum repeater \\ (distributed)} & 3 \& 4 \\
\hline
\end{tabular}
\label{distribution_schemes}
\end{table}

\subsection{Performance of Entanglement Routing}
\label{performance_of_entanglement_routing}

We compare the performance of three representative DER (Q-Cast\cite{shi2020concurrent}, SLMP\cite{pant2019routing}, Greedy\cite{chakraborty2019distributed}) and CER. The performance metrics are fidelity, throughput, entanglement pair consumption, and time cost. In this experiment, these four algorithms are used for the path selection of user A to B in Fig. \ref{example_cellular}.

\begin{figure}[tp]
\centering
\begin{minipage}[t]{0.49\linewidth}
\centering
\includegraphics[width=1.7in]{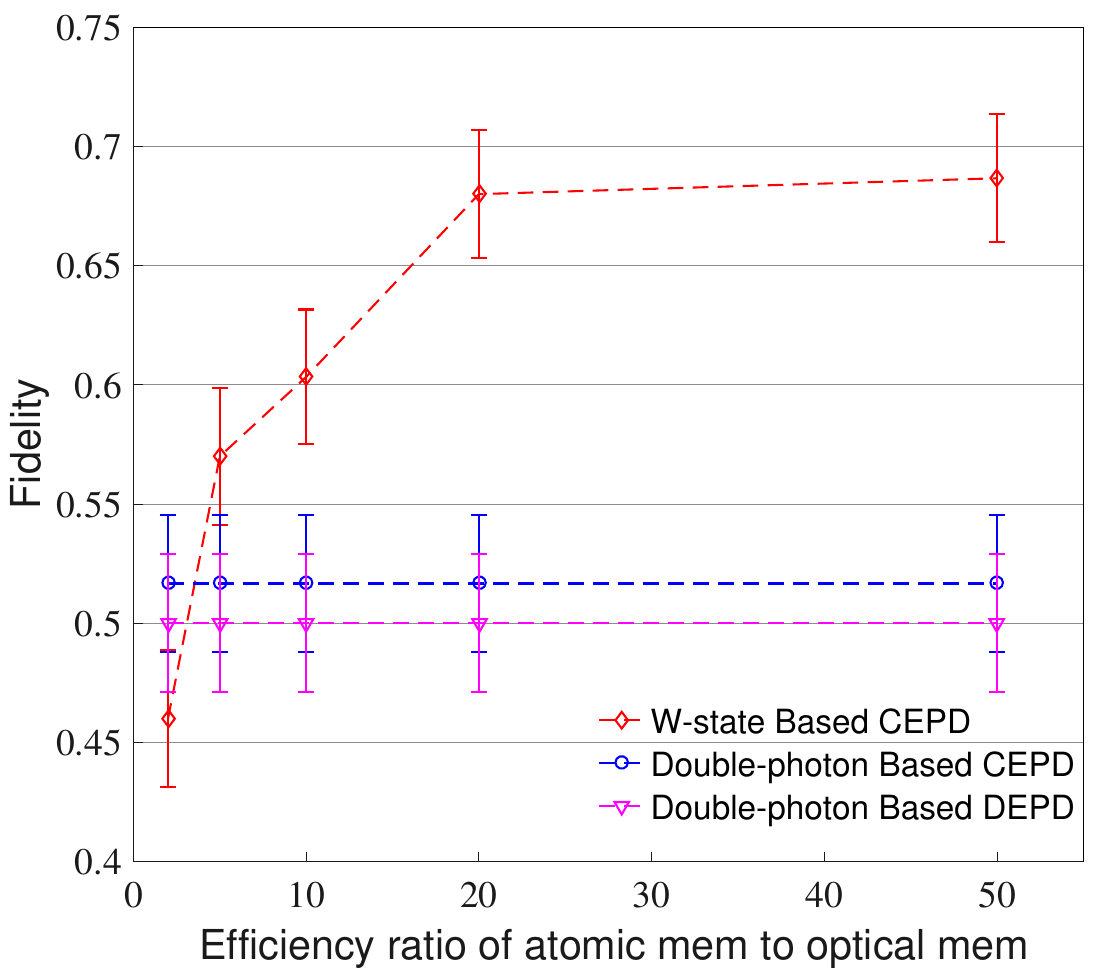}
\caption{Performance of entanglement preparation \& distribution}
\label{entanglement_pd_solution}
\end{minipage}
\begin{minipage}[t]{0.49\linewidth}
\centering
\includegraphics[width=1.8in]{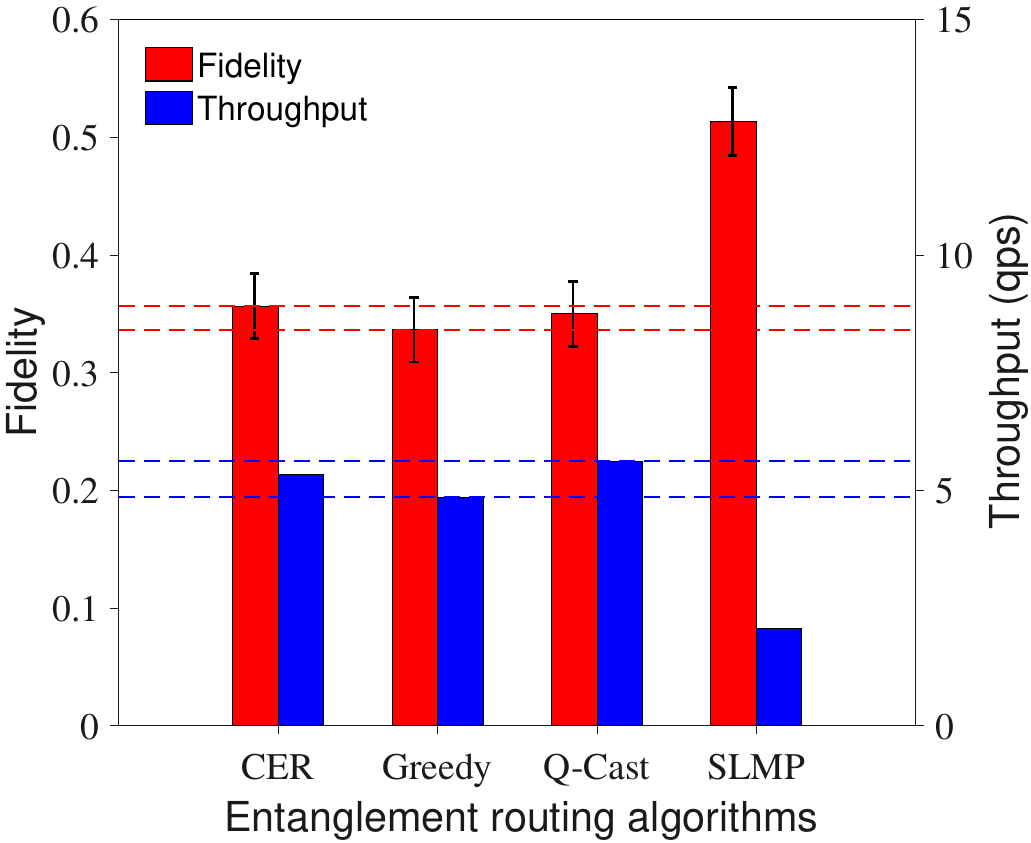}
\caption{Entanglement routing in equivalent parameter network}
\label{routing_equivalent}
\end{minipage}
\end{figure}

{\bfseries Routing in equivalent parameter network:} The equivalent parameter network is an ideal case where every node and link has the same network state. In this experiment, the single channel length is 100 km, the depolarizing rate is set to 0.1, the dephasing rate is set to 0.1, the Q-channel loss init rate is set to 0.01, and the Q-channel loss noise is set to $1.0\times10^{-3}$ dB/km. Fig. \ref{routing_equivalent} shows the performance of four entanglement routing algorithms in the equivalent parameter network. CER, Greedy, and Q-cast have little difference in fidelity and throughput (they are all located in narrow areas delimited by red and blue lines). The reason is that the shortest path is optimal in the equivalent parameter network, and CER, Q-Cast, and Greedy all select the shortest path in this case. However, SLMP has the highest fidelity but the lowest throughput due to its vast time cost and entanglement pair consumption in performing global entanglement distribution in advance to avoid selecting the shortest path with entanglement distribution failure (see Table \ref{routing_cost}). The overhead of SLMP continues to increase with the increasing scale of the quantum Internet, so it is inappropriate to apply SLMP in the global quantum Internet. Based on the experiment result, we can conclude that CER is not inferior to other algorithms in the extreme case of the equivalent parameter network.

\begin{figure}[tp]
\centering
\subfigure[]{
\centering
\includegraphics[width=1.6in]{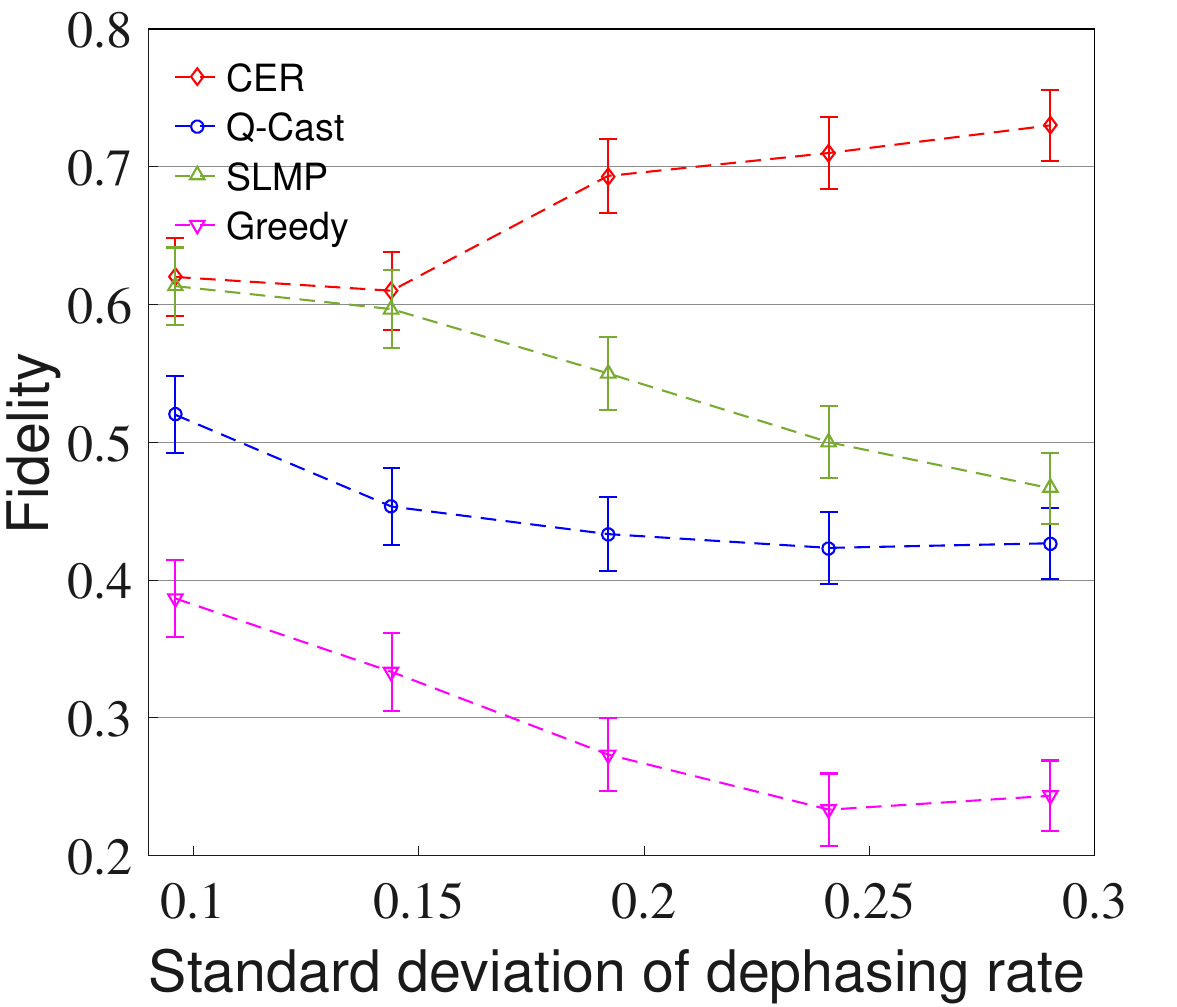}
}
\subfigure[]{
\centering
\includegraphics[width=1.6in]{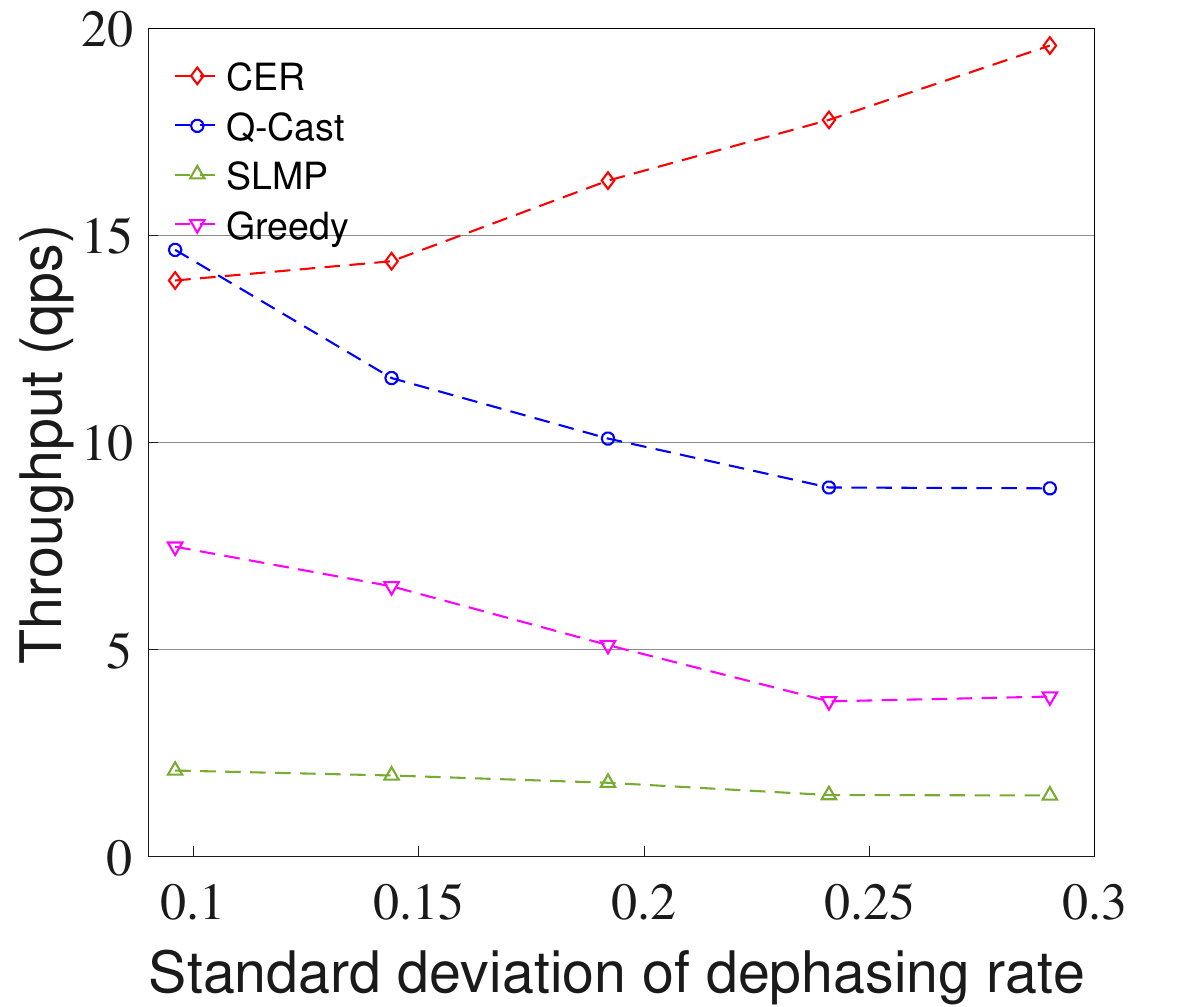}
}
\subfigure[]{
\centering
\includegraphics[width=1.6in]{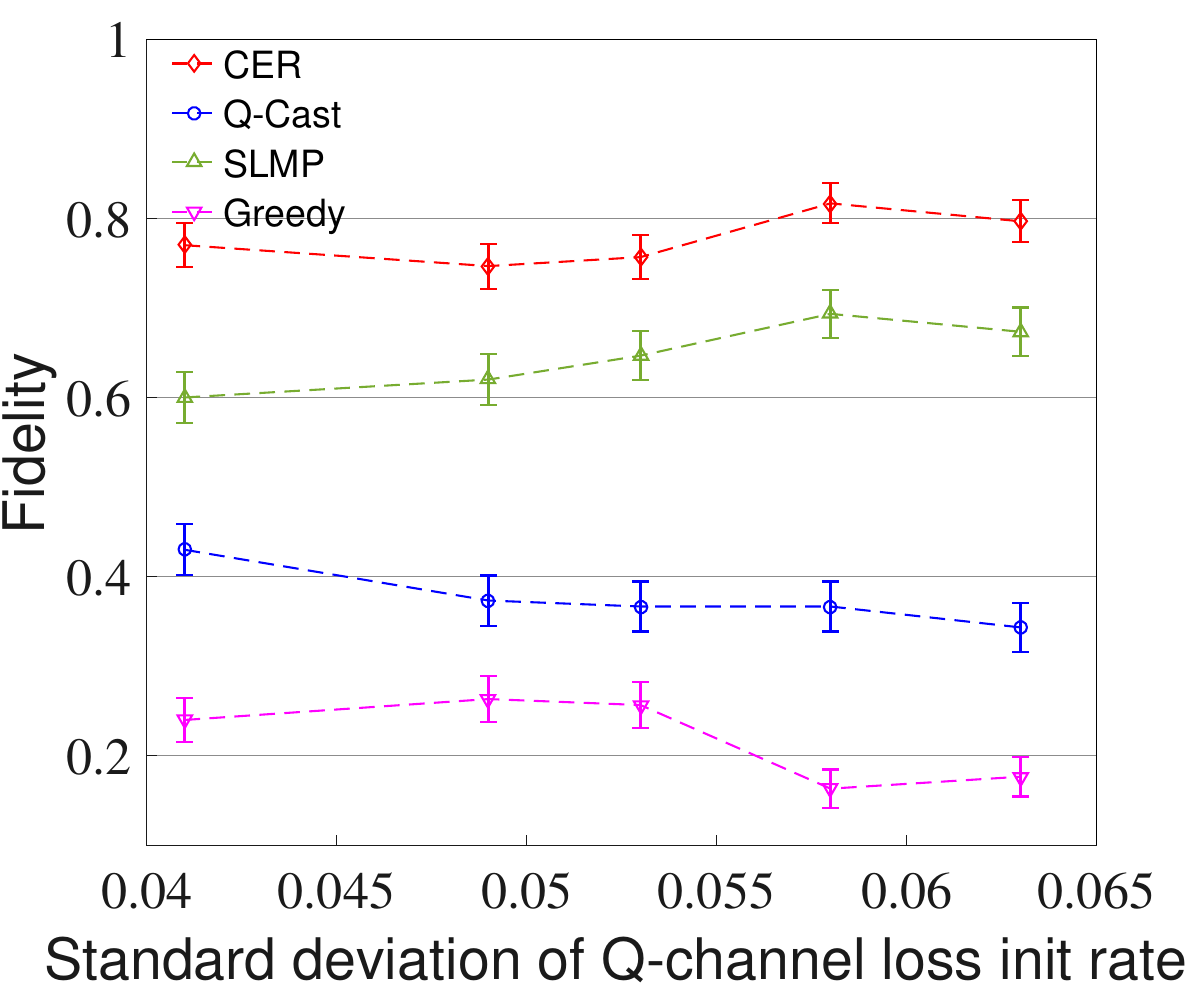}
}
\subfigure[]{
\centering
\includegraphics[width=1.6in]{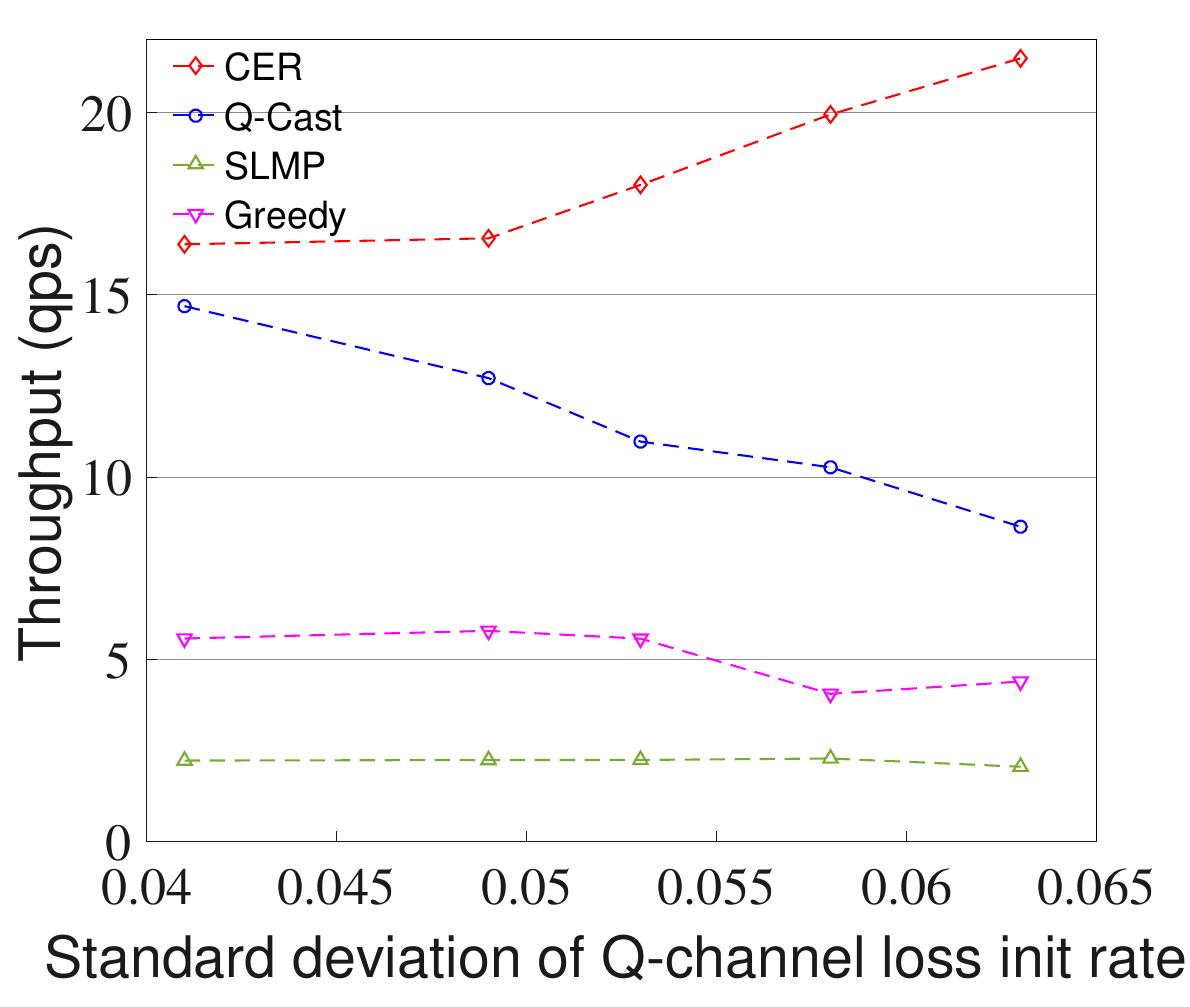}
}
\subfigure[]{
\centering
\includegraphics[width=1.6in]{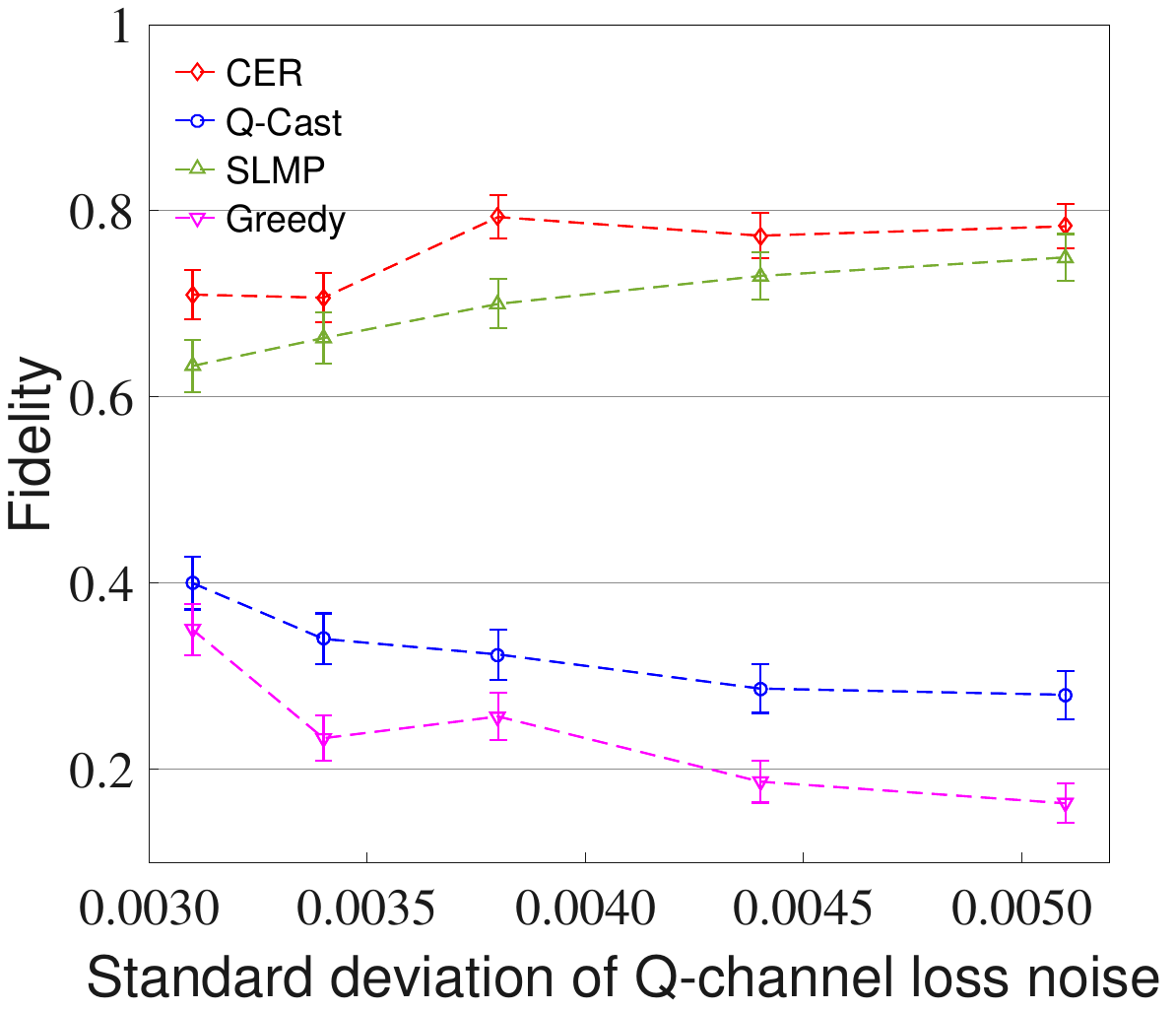}
}
\subfigure[]{
\centering
\includegraphics[width=1.6in]{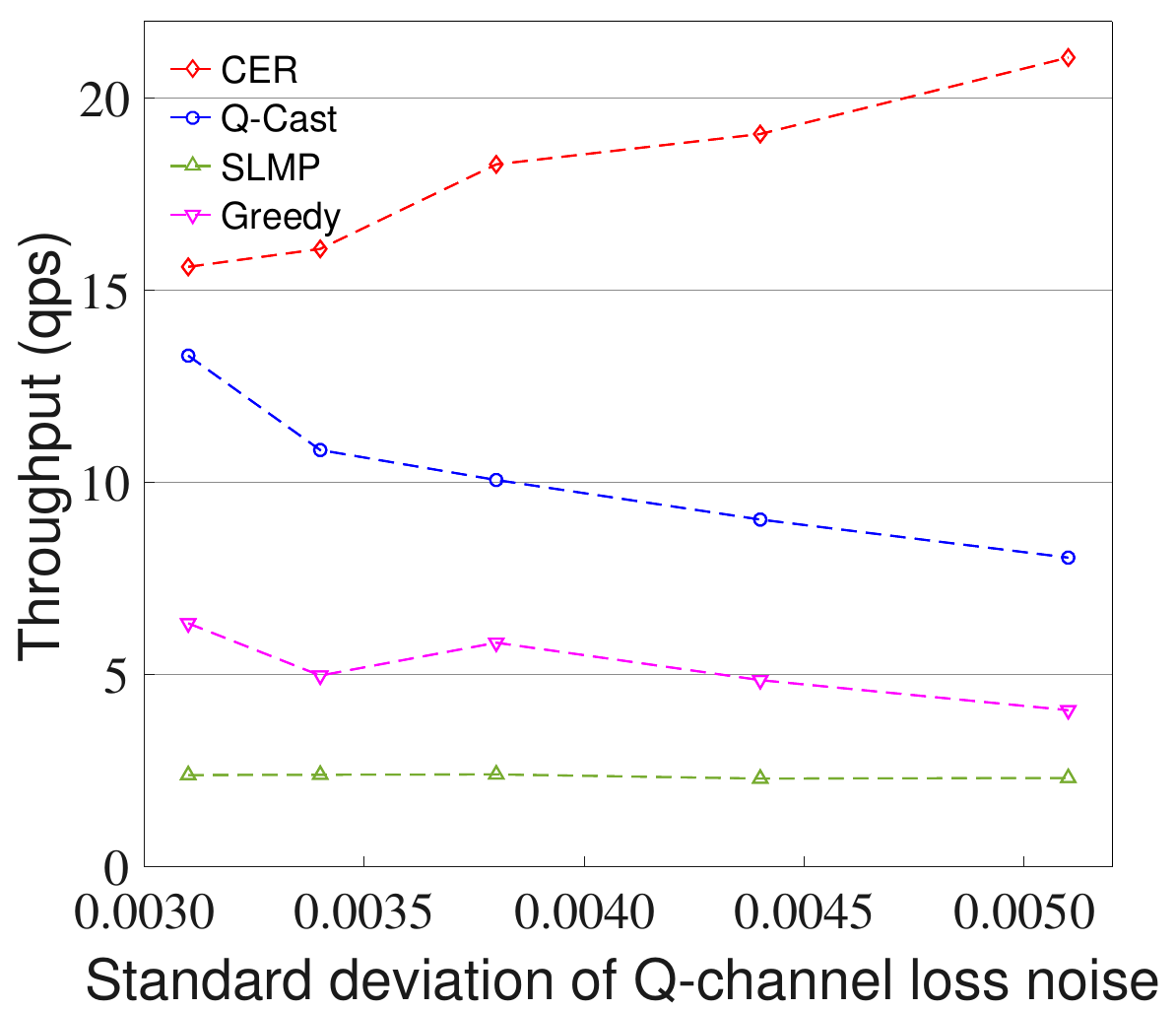}
}
\caption{Entanglement routing in diversified parameter network. Standard deviation represents the difference between devices.}
\label{routing_diversified}
\end{figure}

{\bfseries Routing in diversified parameter network:} We explore the performance of entanglement routing algorithms in the more realistic diversified parameter network where the network states of each node and link are different. We take standard deviations of the environmental parameters to reflect the difference between nodes or links. In particular, we should note that the depolarizing rate is fixed at the same magnitude in this experiment because the depolarizing rate is a relatively stable parameter mainly determined by the memory technology rather than the network load or external environment. In this experiment, the single channel length is 100 km; standard deviations of the dephasing rate are 0.096, 0.144,0.192,0.241,0.290, respectively; standard deviations of the Q-channel loss init rate are 0.041, 0.049, 0.053, 0.058, 0.063, respectively; standard deviations of the Q-channel loss noise are 0.0031, 0.0034, 0.0038, 0.0044, 0.0051, respectively. We should explain that when the standard deviation of one environmental interference changes (e.g., the standard deviation of the dephasing rate), the values of the other two environmental interference must be fixed, guaranteeing the principle of the control variable. Fig. \ref{routing_diversified} shows that the performance of CER is better than other algorithms in a realistic diversified parameter network. Moreover, the difference in fidelity and throughput between CER and the other three algorithms becomes more and more significant when the standard deviation of the dephasing rate, Q-channel loss init rate, and Q-channel loss noise increase. The explanation for this experimental result is that CER utilizes the network state information collected by the controller of hierarchical architecture to make the optimal path. We note that the path evaluation mechanism of CER includes the assessment of environmental interference and hops, and the chosen path has the best fidelity and throughput among all possible paths, so it is the optimal path. In other words, without a hierarchical architecture and global network information, the DER algorithms cannot select the optimal communication path. We also provide the average entanglement pair consumption and time overhead of entanglement routing algorithms, as shown in Table \ref{routing_cost}. Greedy has 5 average pairs consumption and 0.1 ms average time cost because it performs the simplest shortest path routing. Q-Cast has 5.138 average pairs consumption and 0.9 ms average time cost because it enables the recovery path when the primary path fails, resulting in extra entanglement pair consumption and time cost. CER has 6 average pairs consumption and 3.7 ms average time cost because it performs path evaluation and resource reservation to obtain the optimal path. SLMP has 33 average pairs consumption and 132.2 ms average time cost because it performs global entanglement distribution, resulting in huge time and resource costs.

\begin{table}
\centering
\caption{Average entanglement pair consumption and time overhead of entanglement routing algorithms}
\begin{tabular}{|c|c|c|}
\hline  
{\bfseries Algorithms} & {\bfseries Pair consumption} & {\bfseries Time cost}\\
\hline 
Greedy & 5 pairs & 0.1 ms\\
\hline
Q-Cast & 5.138 pairs & 0.9 ms\\
\hline
CER & 6 pairs & 3.7 ms\\
\hline
SLMP & 33 pairs & 132.2 ms\\
\hline
\end{tabular}
\label{routing_cost}
\end{table}

\begin{figure}[tp]
\centering
\subfigure[Fidelity]{
\centering
\includegraphics[width=1.6in]{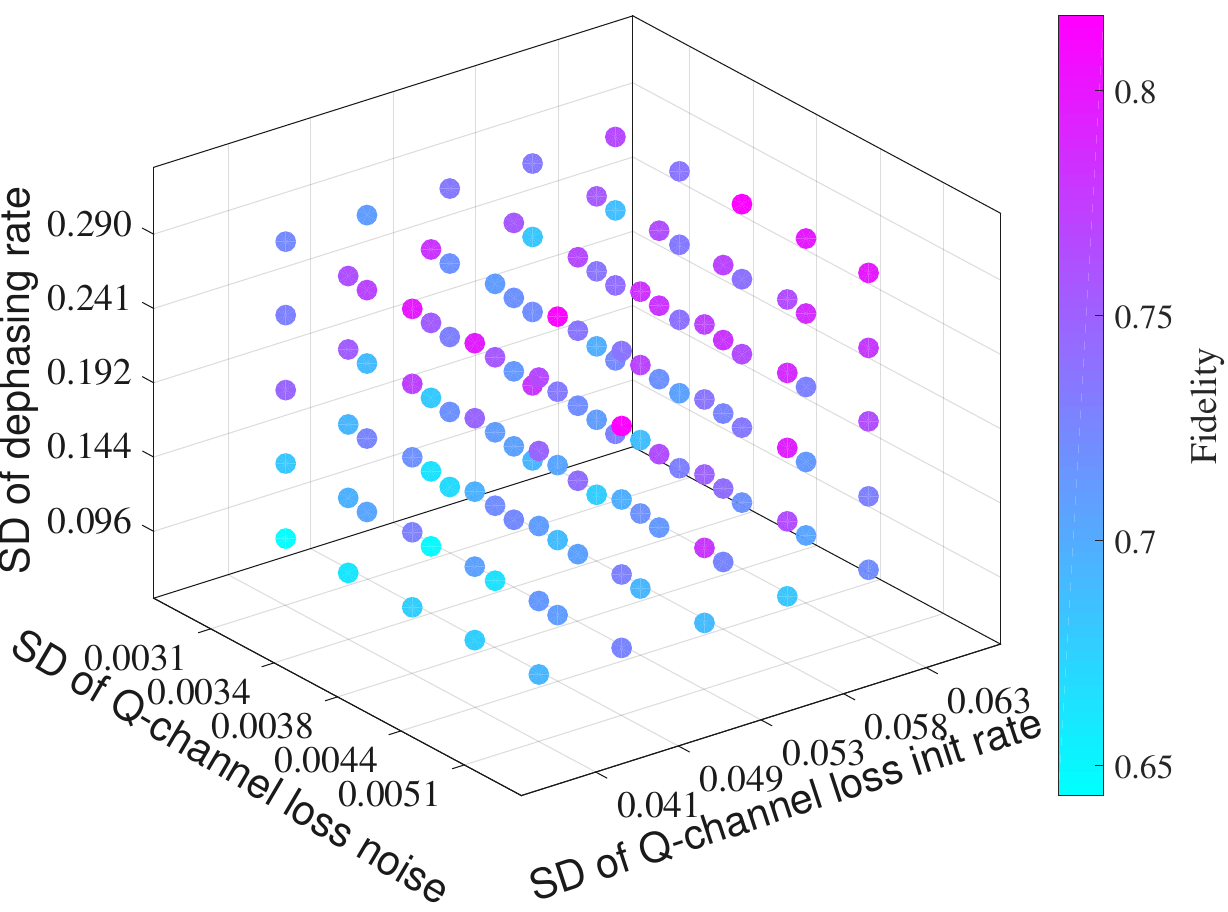}
}
\subfigure[Throughput]{
\centering
\includegraphics[width=1.6in]{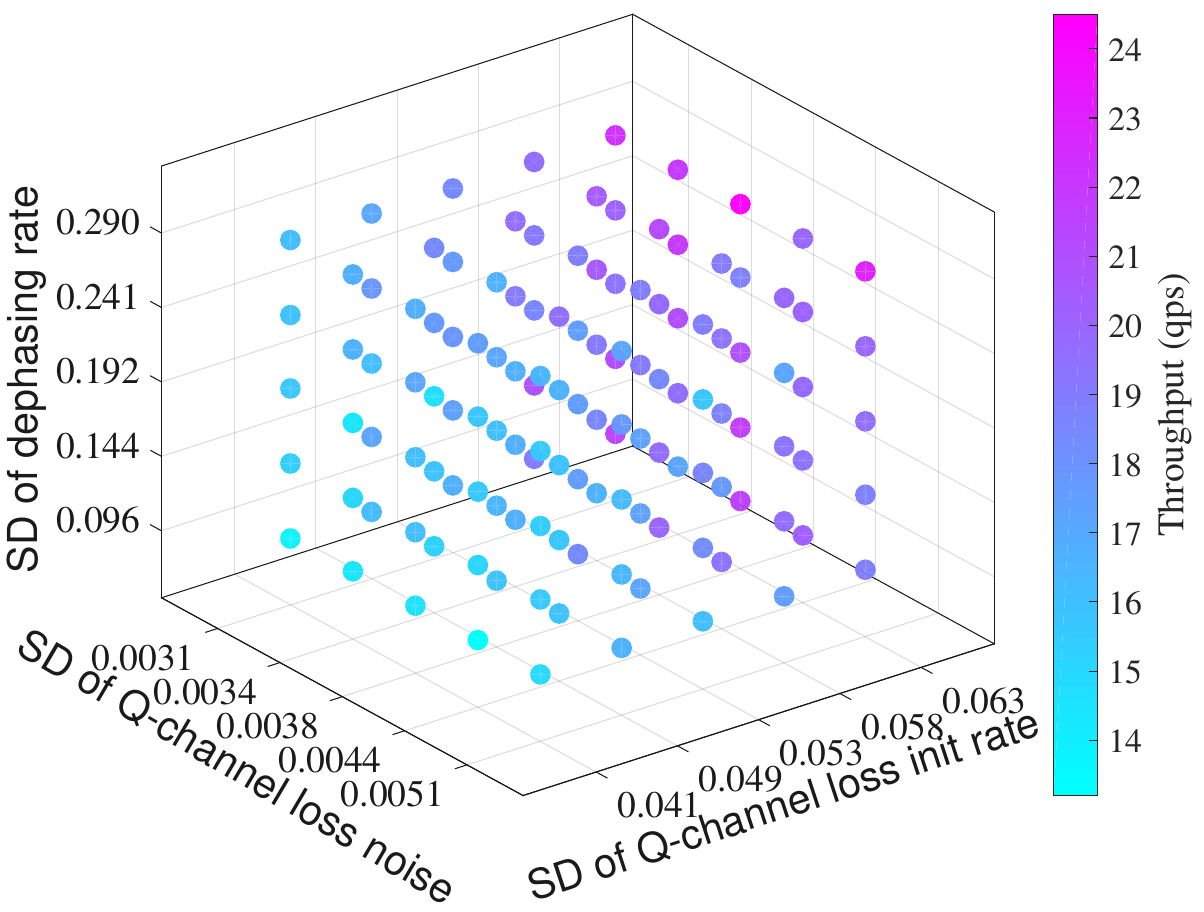}
}
\caption{CER with integrated environmental parameters. The more different the nodes or links in the network, the better the performance of the CER. The CER can choose the optimal path based on the network state.}
\label{CER_integrated}
\end{figure}

{\bfseries CER with integrated environmental parameters:} We integrate the dephasing rate, Q-channel loss init rate, and Q-channel loss noise to observe the performance of CER. The parameters setting in this experiment are consistent with the experiment of the diversified parameter network. However, we combine different environmental interference in this experiment to provide more data and comprehensive results. The 3d-heat-map of Fig. \ref{CER_integrated} shows the results. In Fig. \ref{CER_integrated} (a) and (b), dots closer to the origin are closer to light blue (lower fidelity and throughput), while dots further away from the origin are closer to purple (higher fidelity and throughput). It means that the fidelity and throughput of CER tend to rise with the increase of network states difference. So we can conclude that the more pronounced difference of nodes or links, the more obvious the advantage of CER.

\subsection{Discussion}
From the perspective of entanglement preparation \& distribution, benefiting from the hierarchical architecture that moves entanglement preparation \& distribution from repeaters to local domain controllers, we reduce the number of entanglement preparation facilities and thus reduce the maintenance cost of the quantum Internet. At the same time, a unified entanglement preparator provided by hierarchical architecture enables repeaters to obtain entanglement pairs quickly.

From the perspective of entanglement routing, CER performs optimal routing with the help of the global vision of hierarchical architecture. In contrast, DER algorithms cannot perform optimal routing because distributed architecture does not provide the global network state.

In conclusion, compared with distributed architecture, hierarchical architecture has three advantages: 1) reducing maintenance costs by moving the entanglement preparation \& distribution from massive quantum repeaters to fewer local domain controllers; 2) increasing the efficiency by centralized entanglement distribution with the local domain controller; 3) making optimal routing possible by global network state collection. Although the maximum performance improvement for entanglement distribution is 37.3\%, the hierarchical architecture has great potential based on advances in atomic memory technology\cite{heshami2016quantum}, scalability, and optimal routing support.

\section{Conclusion and Future Work}
\label{conclusion}
This paper proposes hierarchical quantum Internet architecture for the first time, which provides lower maintenance cost, efficient entanglement distribution, and optimal entanglement routing. The quantum Internet is still in its infancy, and this work provides a new perspective for early quantum Internet architecture research. Even if we still have a long way to the quantum Internet, the hierarchical architecture can be a potential solution. We hope this work can motivate a new potential research area and attract more computer science and physics researchers to cooperate on the quantum Internet.

Here, we identify two possible future research topics related to our proposed hierarchical quantum Internet architecture: 1) research on cooperative mechanisms between multiple central controllers and 2) integrating the existing quantum repeater technology and entanglement purification into the hierarchical architecture.

{\appendix
\subsection{Central State Matrix and Local State Matrix}
\label{example_LSM_CSM}
The Central State Matrix (CSM) and Local State Matrix (LSM) are located on the central controller and local domain controller, respectively. LSMs report local domain information to CSM via classical channels. They are implemented by Tree Struct (see source code in\cite{hierarchicalquantumarch2023}) and aim to collect global network states, including:

\par{\bfseries 1) DomainName}: It represents the name of the domain.

{\bfseries 2) DeviceName:} It represents the name of the device.

{\bfseries 3) DeviceState:} It represents the state of the device. The state can be ``normal'' or ``maintain''. ``maintain'' means the device is unavailable. When a device fails to complete entanglement preparation or distribution several times, its DeviceState is set to ``maintain" and then checked by the network administrator. DeviceState is used for entanglement routing to prevent broken devices from appearing on the path.

{\bfseries 4) MemoryName:} It represents the name of quantum memory.

{\bfseries 5) MemoryState:} It represents the state of quantum memory. The state can be ``idle'' or ``occupy''.

{\bfseries 6) AimPair:} It represents the quantum memory name of the entanglement pair stored. AimPair will be updated after the entanglement distribution.

{\bfseries 7) AimCommunication:} It represents the corresponding communication of quantum memory. AimComunication will be updated after the entanglement distribution.

{\bfseries 8) AimNode:} It represents the target quantum device of the local domain controller's entanglement preparation \& distribution. AimNode will be updated after the entanglement preparation.

{\bfseries 9) DistributionState:} It represents the state of entanglement distribution in the user or repeater.

{\bfseries 10) PreparationState:} It represents the state of entanglement preparation in the local domain controller.

{\bfseries 11) SwappingState:} It represents the state of entanglement swapping in the repeater.

{\bfseries 12) TeleportationState:} It represents the state of quantum teleportation in the user.

{\bfseries 13) LinkState:} It represents the entanglement preparation \& distribution success rate. When the quantum entanglement source is stable, the success rate of entanglement preparation \& distribution mainly depends on the quality of the quantum channel, so it is called LinkState. LinkState increases according to statistics only if both entanglement preparation and entanglement distribution are successful; otherwise, LinkState decreases. LinkState is one basis for entanglement routing to select the optimal path.

{\bfseries 14) SwappingSuccessRate:} It represents the entanglement swapping success rate. We obtain the SwappingSuccessRate according to the statistics.

\subsection{Double-photon Based Centralized Entanglement Preparation \& Distribution}
\label{double_photon_CEPD}
Here, we propose the Double-photon Based CEPD as an experimental comparison to W-state Based CEPD.

As shown in Fig. \ref{double_photon_based_CEPD}, Double-photon Based CEPD requires the local domain controller to prepare photon-photon entanglement and distribute them directly to optical quantum memories of repeaters. Double-photon Based CEPD does not require extra operations and photons transmission, but the qubit lifetime is short due to the optical quantum memory.

\subsection{The Impact Comparison Between Quantum Channel Quality and Dephasing Rate}
\label{channel_quality_and_dephasing}
As shown in Fig. \ref{noise_limitation}, when the noise of a 100 km channel is greater than or equal to 0.2 dB/km, the success rate of entanglement distribution is zero, so the maximum acceptable noise of a 100 km channel is 0.2 dB/km, and we can quantify the channel quality by formula \ref{channel_quality}.

\begin{equation}
\begin{aligned}
ChannelQuality = &1 - (\text{Q-channel loss init rate}  + \\ & \text{Q-channel loss noise} \div 0.2) \div 2
\end{aligned}
\label{channel_quality}
\end{equation}

The experiment result is shown in Table \ref{comparison_c_d}. When the channel quality drops from 1 (Q-channel loss init rate = 0, Q-channel loss noise = 0 dB/km) to 0.85 (Q-channel loss init rate = 0.2, Q-channel loss noise = 0.02 dB/km), the fidelity variation is 1; while when the dephasing rate increases from 0 to 0.15 (the quantum operation success rate drops from 1 to 0.85), the fidelity variation is 0.5. Thus, we can conclude that the impact on the communication of quantum channel quality is about twice than dephasing rate. 

\begin{table}[h]
\centering
\caption{Impact of channel quality and dephasing rate}
\begin{tabular}{|c|c|c|c|c|}
\hline  
\multicolumn{1}{|c|}{\multirow {2}{*}{}} & \multicolumn{2}{c|}{\textbf {Channel quality}} & \multicolumn{2}{c|}{\textbf {Dephasing rate}}\\
\cline{2-5}
\multicolumn{1}{|c|}{} & 1 & 0.85 & 0 & 0.15 \\
\hline 
\textbf {Fidelity} & 1 & $\approx0$ & 1 & $\approx0.5$ \\
\hline
\makecell[c]{\textbf {Fidelity variation}} & \multicolumn{2}{c|}{$\approx1$} &\multicolumn{2}{c|}{$\approx0.5$}\\
\hline
\end{tabular}
\label{comparison_c_d}
\end{table}

\subsection{The Importance of Environmental Interference in Entanglement Routing}
\label {environmental_interference_importance}
We set three paths with different hops (4-hop short path A, 5-hop long path B, and 4-hop short path C) and three groups of environmental parameters. The order of environmental parameter size in each group is path B $<$ path A $<$ path C. From group 1 to group 3, the environmental parameter difference between path B and paths A, C gradually increases. The setting means that we assign lower environmental interference to the long path than to the short paths. 

Bars in Fig. \ref{fidelity_integrated_conditions} show the communication fidelity of three paths in different groups. The fidelity rankings in the three groups are all long path B $>$ short path A $>$ short path C. Lines in Fig. \ref{fidelity_integrated_conditions} show the fidelity difference between long path B and short paths A, C in the three groups. The fidelity difference increases with the increase of environmental parameter difference. The result demonstrates the powerful influence of environmental interference and the importance of collecting network states in optimal entanglement routing.

\begin{figure} [tp]
\centering
\includegraphics[width=3in]{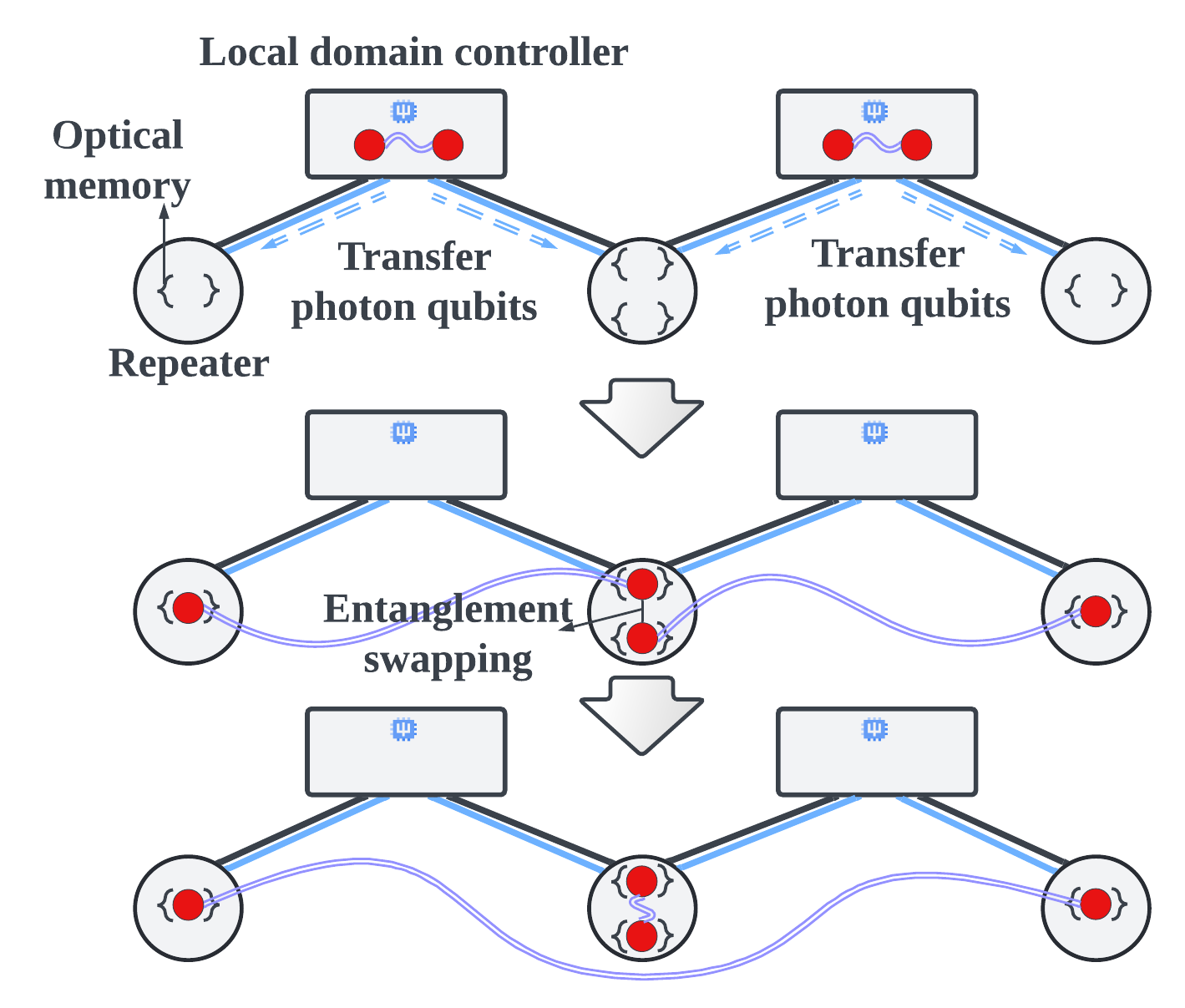}
\caption{Double-photon Based Centralized Entanglement Preparation \& Distribution. The local domain controller prepares and directly distributes photon-photon entanglement without extra operations and photons transmission.}
\label{double_photon_based_CEPD}
\end{figure}

\begin{figure}[tp]
\begin{minipage}[t]{0.49\linewidth}
\centering
\includegraphics[width=1.8in]{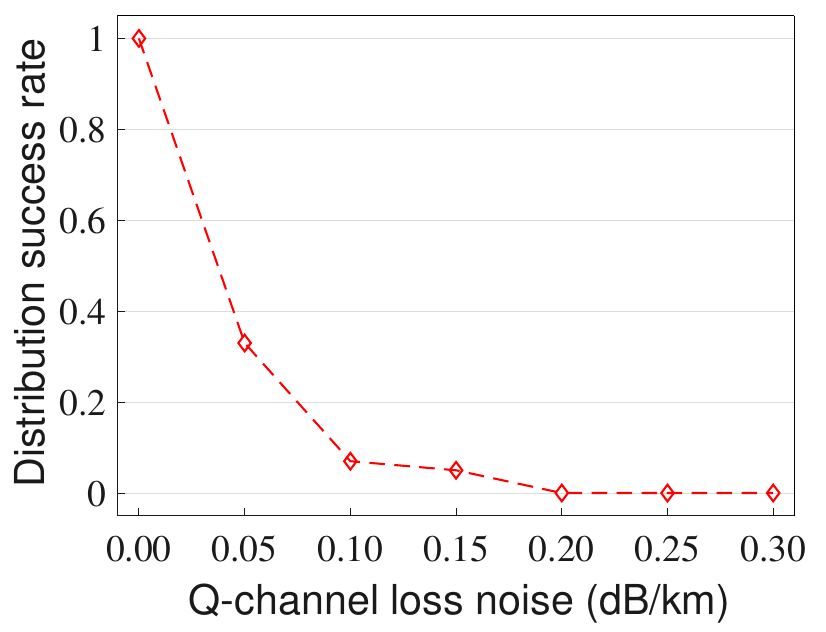}
\caption{Noise limitation. The quantum channel length is 100 km.}
\label{noise_limitation}
\end{minipage}
\begin{minipage}[t]{0.49\linewidth}
\centering
\includegraphics[width=1.8in]{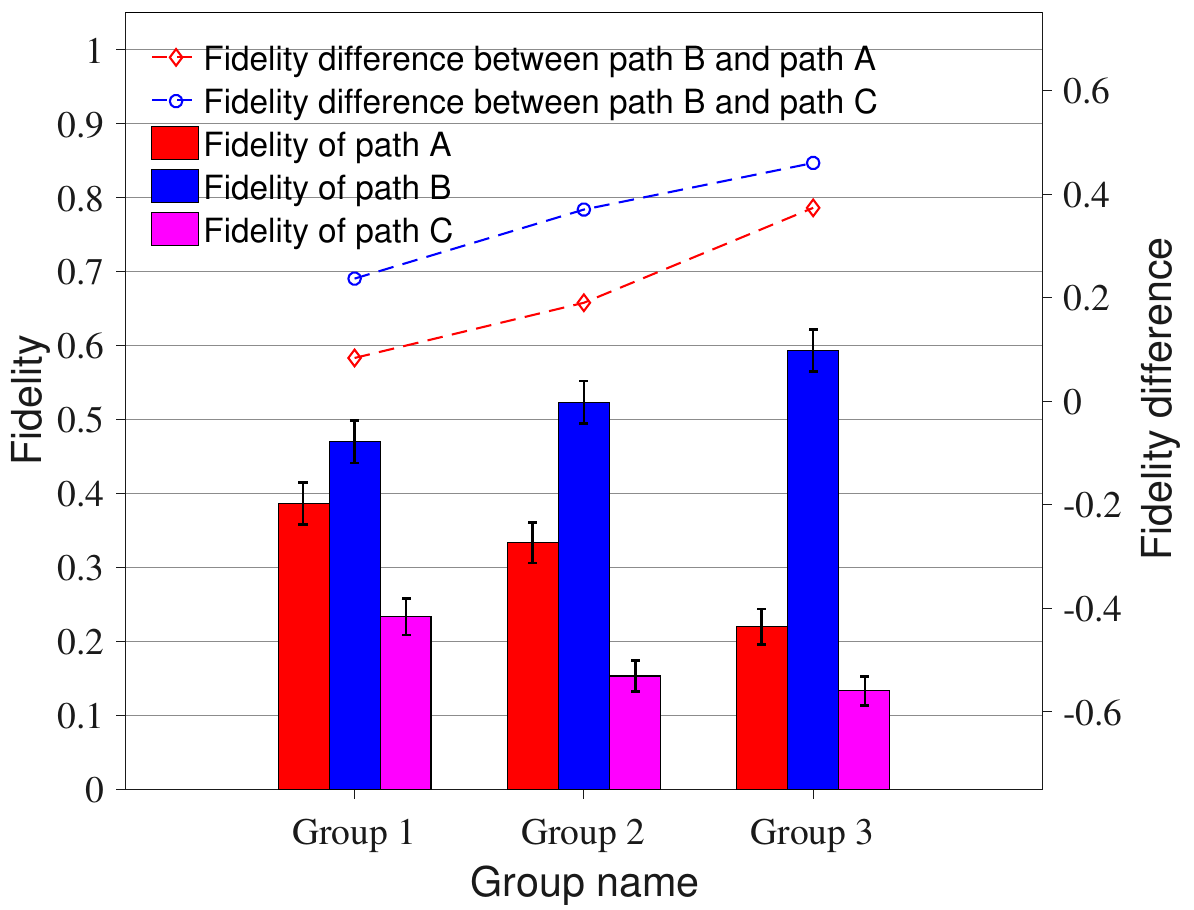}
\caption{The importance of environmental interference in entanglement routing. Long path B has lower environmental interference than short paths A and C. The difference in environmental interference between paths gradually increases from Group 1 to 3.}
\label{fidelity_integrated_conditions}
\end{minipage}
\end{figure}

\subsection{Simulation of the W-state Based Centralized Entanglement Preparation \& Distribution}
\label{simulation_W_state}
Because the NetSquid\cite{netsquid2021} does not support atom-photon-photon W-state preparation, we adjust the environmental parameters to simulate the W-state Based CEPD.

Firstly, W-state Based CEPD utilizes atomic memory to store qubits, which is more efficient than optical memory\cite{heshami2016quantum}. Therefore, we simulate atomic memory by decreasing the depolarizing rate. Assuming the efficiency of atomic memory is N times that of optimal memory, we can get the depolarizing rate of atomic memory according to formula \ref{atomic_depolarizing}. AMDR is for Atomic Memory Depolarizing Rate, and OMDR is for Optimal Memory Depolarizing Rate.

\begin{equation}
AMDR = \frac{OMDR}{N}
\label{atomic_depolarizing}
\end{equation}

Secondly, W-state Based CEPD requires five quantum operations for each distribution. Thus, For N-hop communication, W-state Based CEPD requires $(N + 1) \times 5 + N$ quantum operations, while Double-photon Based CEPD requires only N quantum operations. We can get the ratio of quantum operations of W-state Based CEPD to that of Double-photon Based CEPD, as shown in formula \ref{ratio_operation}. We reflect the extra quantum operations by raising the dephasing rate. The experiment in Sec. \ref{performance_of_entanglement_pd} is 4-hop communication, so the ratio of quantum operations is 7.25 (round it up to 8). Formula \ref{wstate_dephasing} shows the dephasing rate of W-state Based CEPD. WDR is for W-state Based CEPD Dephasing Rate, and DPDR is for Double-photon Based CEPD Dephasing Rate. The $(1-DPDR)^8$ in formula \ref{wstate_dephasing} means the probability of all eight quantum operations successfully performing.

\begin{equation}
6<6+\frac{5}{N} \leq11(N=1,2,3,4...)
\label{ratio_operation}
\end{equation}

\begin{equation}
WDR =1-(1-DPDR)^8
\label{wstate_dephasing}
\end{equation}

Thirdly, for N-hop communication, W-state Based CEPD needs to transmit $4 \times (N + 1)$ photons, while Double-photon Based CEPD only requires $2 \times (N + 1)$ photons. So the ratio of transmitted photons is 2. We reflect the extra transmitted photons of W-state Based CEPD by raising the Q-channel loss init rate and Q-channel loss noise. Formulas \ref{wstate_loss_init} and \ref{wstate_loss_noise} show the Q-channel loss init rate and Q-channel loss noise of W-state Based CEPD. WLIR is for W-state Based CEPD Q-channel Loss Init Rate, DPLIR is for Double-photon Based CEPD Q-channel Loss Init Rate, WLN is for W-state Based CEPD Q-channel Loss Noise, and DPLN is for Double-photon Based CEPD Q-channel Loss Noise.

\begin{equation}
WLIR = 1-(1-DPLIR)^2
\label{wstate_loss_init}
\end{equation}

\begin{equation}
WLN = DPLN \times 2
\label{wstate_loss_noise}
\end{equation}
}

\bibliographystyle{IEEEtran}
\def\bibfont{\small}
\bibliography{reference}

\begin{IEEEbiography}[{\includegraphics[width=1in, height=1.25in,clip,keepaspectratio]{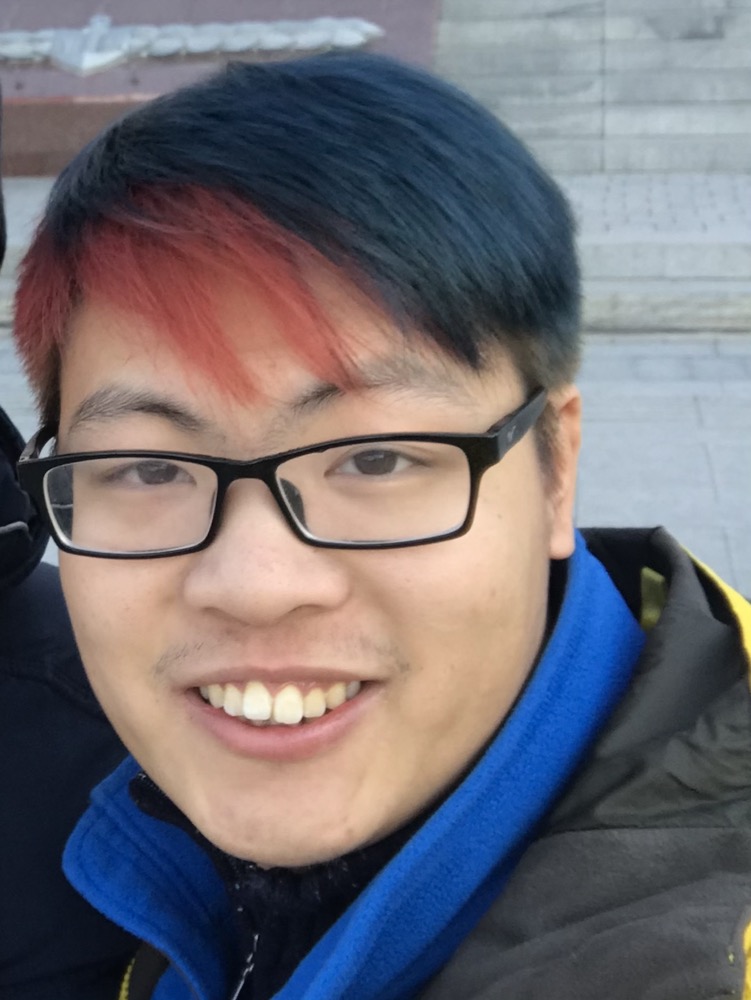}}]{Binjie He}
(Student Member, IEEE) received the B.S. and M.S. degrees from Fuzhou University, China, in 2016 and 2019, respectively. He worked for Cisco as a software engineer from 2019 to 2021 and designed a monitoring platform architecture for global data centers. He is currently pursuing his Ph.D. degree at the College of Computer and Data Science, Fuzhou University. He is the runner-up of the Wilkes Award in 2021. His research area includes quantum Internet, software defined networks, flow monitoring, load balancing algorithms, and artificial intelligence.
\end{IEEEbiography}

\begin{IEEEbiography}[{\includegraphics[width=1in, height=1.25in,clip,keepaspectratio]{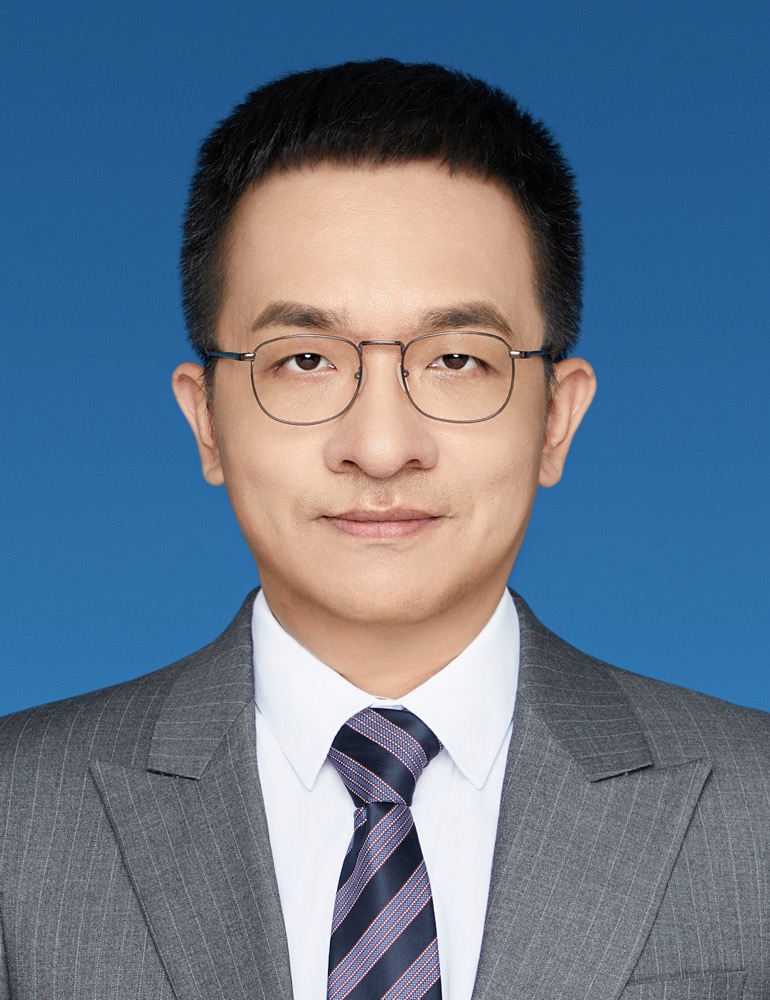}}]{Dong Zhang}
(Member, IEEE) received the B.S. and Ph.D. degrees from Zhejiang University, China, in 2005 and 2010, respectively. He visited Alabama University, USA, as a Visiting Scholar, from 2018 to 2019. He is currently a Professor with the College of Computer and Data Science, Fuzhou University, China. His research interests include software defined networking, network virtualization, and Internet QoS.
\end{IEEEbiography}

\begin{IEEEbiography}[{\includegraphics[width=1in, height=1.25in,clip,keepaspectratio]{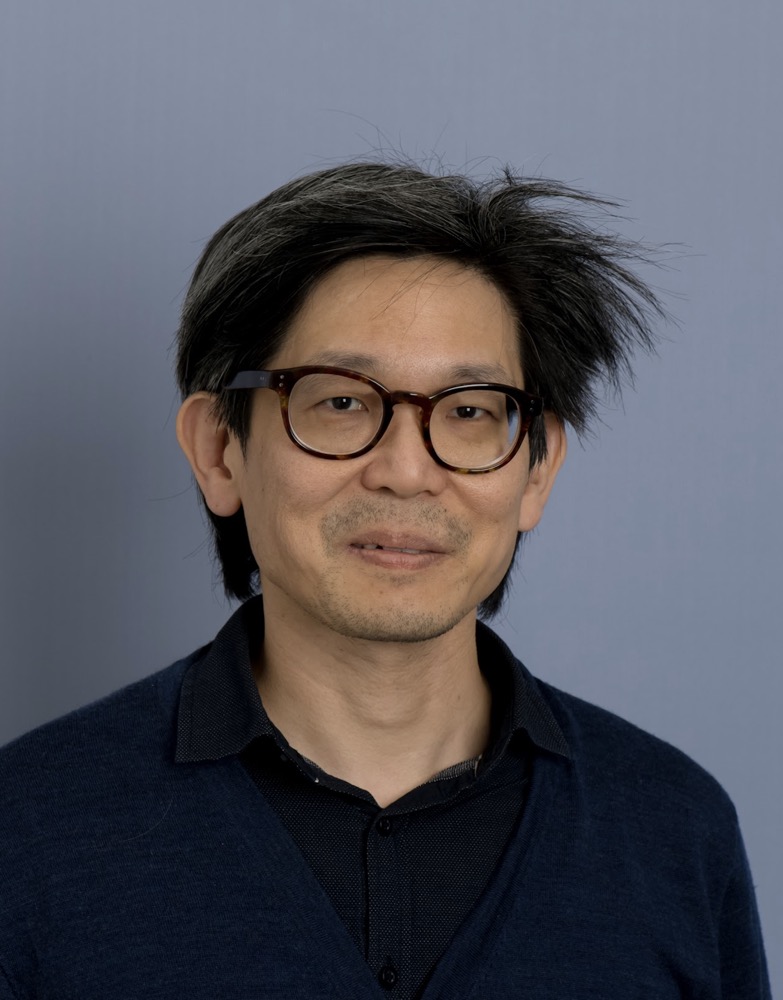}}]{Seng W. Loke}
(Member, IEEE), Ph.D. (1998), is a Professor in Computer Science for the School of Information Technology, Deakin University, Australia, as well as Director of the Centre for Software, Systems and Society (CS3) and Co-Director of the Internet of Things Platforms and Applications Lab in the Centre for Internet of Things Ecosystem Research and Experimentation (CITECORE). His research interests are mainly in the Internet of Things, including quantum Internet computing, context-aware computing, mobile and pervasive computing.
\end{IEEEbiography}

\begin{IEEEbiography}[{\includegraphics[width=1in, height=1.25in,clip,keepaspectratio]{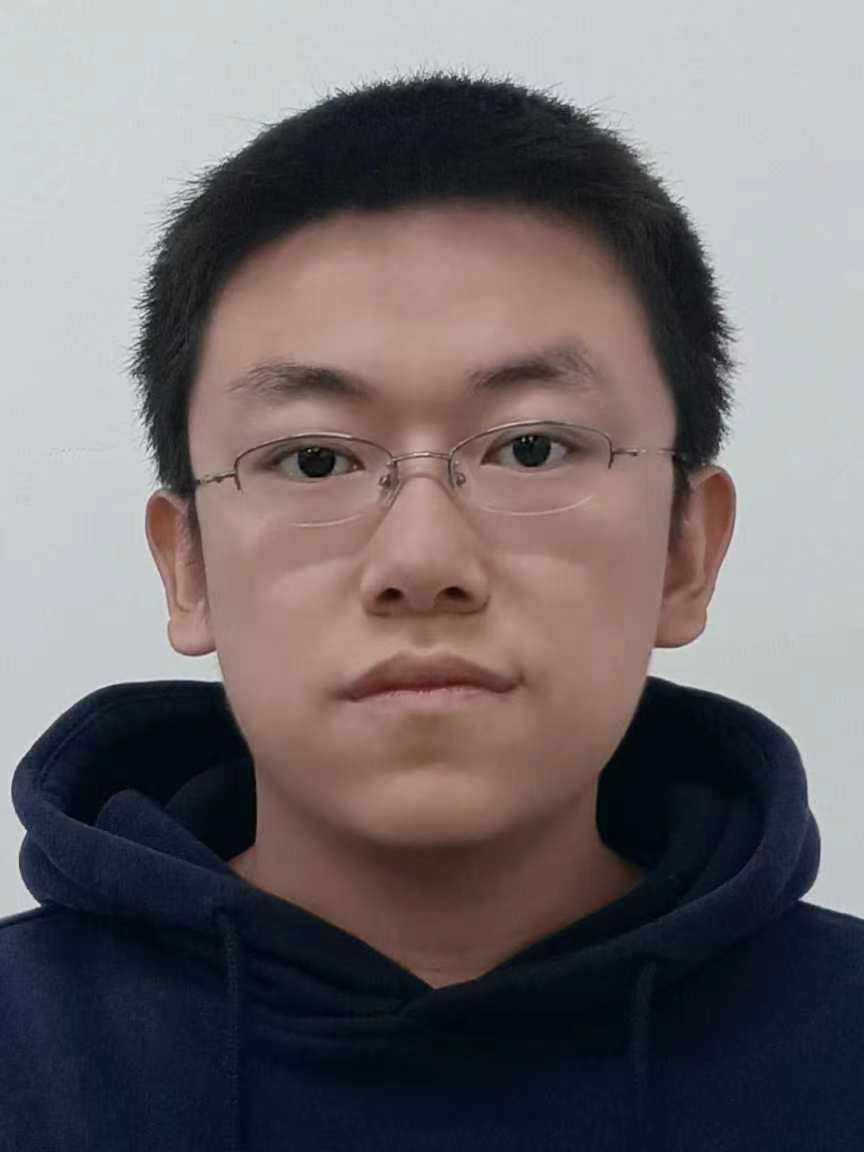}}]{Shengrui Lin}
received the B.S. degree in Xi'an University of Science and Technology in 2019. He is currently pursuing his Ph.D. degree at the College of Computer and Data Science, Fuzhou University. His current research interests include programmable networks and in-network intelligence. 
\end{IEEEbiography}

\begin{IEEEbiography}[{\includegraphics[width=1in, height=1.25in,clip,keepaspectratio]{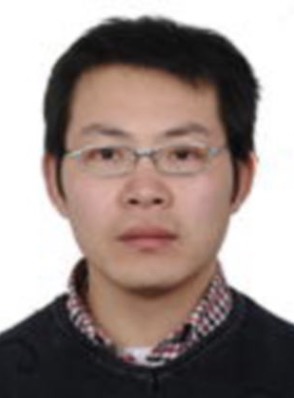}}]{Luke Lu}
(Senior Tech Lead, Cisco Systems) received the B.S. and M.S. degrees from Hohai University, China, in 2008, focusing on computer networks and software engineering. He has worked for Cisco as a senior tech leader in data center networks for decades. He achieved the CCIE Security certification in 2015. Currently, he is passionately engaged in research at the intersection of artificial intelligence and data center networks, aiming to enhance efficiency and security in these critical infrastructures.
\end{IEEEbiography}

\end{document}